\documentclass[12pt,draftcls,onecolumn,journal]{IEEEtran} 
\usepackage{times}
\usepackage{graphicx,subfigure}
\usepackage{amsmath,amsfonts}
\usepackage{float}

\newtheorem{thm}{Theorem}
\newtheorem{lem}{Lemma}

\begin{document}

\title{The Three Node Wireless Network: Achievable Rates and Cooperation Strategies}

\author{Lifeng~Lai,~Ke~Liu,~and~Hesham~El~Gamal\thanks{The authors are with the ECE Department at the Ohio State University.
This work was funded in part by the National Science Foundation
under Grants CCR~0118859, ITR~0219892, and
CAREER~0346887.}\\\{lail,liuk,helgamal\}@ece.osu.edu} \maketitle

\begin{abstract}
We consider a wireless network composed of three nodes and limited
by the half-duplex and total power constraints. This formulation
encompasses many of the special cases studied in the literature
and allows for capturing the common features shared by them. Here,
we focus on three special cases, namely 1) Relay Channel, 2)
Multicast Channel, and 3) Conference Channel. These special cases
are judicially chosen to reflect varying degrees of complexity
while highlighting the common ground shared by the different
variants of the three node wireless network. For the relay
channel, we propose a new cooperation scheme that exploits the
wireless feedback gain. This scheme combines the benefits of
decode-and-forward and compress-and-forward strategies and avoids
the idealistic feedback assumption adopted in earlier works. Our
analysis of the achievable rate of this scheme reveals the
diminishing feedback gain at both the low and high signal-to-noise
ratio regimes. Inspired by the proposed feedback strategy, we
identify a greedy cooperation framework applicable to both the
multicast and conference channels. Our performance analysis
reveals several {\em nice} properties of the proposed greedy
approach and the central role of cooperative source-channel coding
in exploiting the receiver side information in the wireless
network setting. Our proofs for the cooperative multicast with
side-information rely on novel nested and independent binning
encoders along with a list decoder.
\end{abstract}

\section{Introduction}
We are in the midst of a new wireless revolution, brought on by
the adoption of wireless networks for consumer, military,
scientific, and wireless applications. For example, the consumer
potential is clearly evident in the exploding popularity of
wireless LANs and Bluetooth-protocol devices. The military
potential is also clear: wireless networks can be rapidly
deployed, and the failure of individual nodes does not imply the
failure of the network. Scientific data-collection applications
using wireless sensor networks are also gaining in numbers. These
applications have sparked a renewed interest in network
information theory. Despite the recent progress ( see
\cite{Xie-Kumar-scaling,Goldsmith.capacity,Gupta.capacity,Tse.mobility,
Kramer.relay} and references wherein), developing a unified theory
for network information flow remains an elusive task.

In our work, we consider, perhaps, the most simplified scenario of
wireless networks. Our network is composed of only three nodes and
limited by the half-duplex and total power constrains. Despite
this simplicity, this model encompasses many of the special cases
that have been extensively studied in the literature. These
special {\em channels}\footnote{With a slight abuse of notation,
we interchange ``channel'' and ``network'' in different places of
the sequel for maximal consistency with the literature.} are
induced by the traffic generated at the nodes and the requirements
imposed on the network\footnote{For example, the relay channel
corresponds to the special case where the traffic is generated at
one node and is required to be transmitted to only one of the
remaining two nodes.}. More importantly, this model exposes the
common features shared by these special cases and allows for
constructing universal cooperation strategies that yield
significant performance gains. In particular, we focus here on
three special cases, namely 1) Relay Channel, 2) Multicast
Channel, and 3) Conference Channel. These channels are defined
rigorously in Section~\ref{sec:general}. We adopt a greedy
framework for designing cooperation strategies and characterize
the achievable rates of the proposed schemes. Our analysis reveals
the structural similarities of the proposed strategies, in the
three special cases, and establishes the asymptotic optimality of
such strategies in several cases. More specifically, our
contributions can be summarized as follows.

\begin{enumerate}
\item We propose a novel cooperation strategy for the relay
channel with feedback. Our scheme combines the benefits of both
the decode-and-forward (DF) and compress-and-forward (CF)
strategies and avoids the idealistic assumptions adopted in
earlier works. Our analysis of the achievable rate of the proposed
strategy reveals the diminishing gain of feedback in the
asymptotic scenarios of low and large signal-to-noise ratio. We
further establish the sub-optimality of orthogonal cooperation
strategies ( \cite{Gamal.relay,Laneman.halfduplex}) in this {\bf
half duplex} setting. \item Inspired by the feedback strategy for
the relay channel, we construct a greedy cooperation strategy for
the multicast scenario. Motivated by a greedy approach, we show
that the {\em weak} receiver is led to help the {\em strong}
receiver first\footnote{The notions of weak and strong receivers
will be defined rigorously in the sequel.}. Based on the same
greedy motivation, the strong user starts to assist the weak
receiver after successfully decoding the transmitted codeword. We
compute the corresponding achievable rate achieved by and use it
to establish the significant gains offered by this strategy, as
compared with the non-cooperative scenario.  \item Motivated by
the sensor networks application, we identify the conference
channel model as a special case of our general formulation. In
this model, the three nodes observe correlated date streams and
every node wishes to communicate its observations to the other two
nodes. Our proposed cooperation strategy in this scenario consists
of three stages of {\em multicast with side information}, where
the multicasting order is determined by a low complexity greedy
scheduler. In every stage, we use a cooperation strategy obtained
as a generalization of the greedy multicast approach. This
strategy highlights the central role of cooperative source-channel
coding in exploiting the side information available at the
receivers. By contrasting the minimum energy required by the
proposed strategy with the genie-aided and non-cooperative
schemes, we establish its superior performance. \item We identify
the greedy principle as the basis for constructing efficient
cooperation strategies in the three considered scenarios. Careful
consideration of other variants of the three node network reveals
the fact that such principle carries over with slight
modifications.
\end{enumerate}

The rest of the paper is organized as follows.
Section~\ref{sec:general} introduces our modelling assumptions and
notation. In Section~\ref{sec:relay}, we present the new
cooperation strategy for the wireless relay channel with {\em
realistic} feedback and analyze its performance. Building on the
relay channel strategy, Section~\ref{sec:multicast} develops the
greedy cooperation framework for the multi-cast channel. We devote
Section~\ref{sec:conference} to the conference channel. Finally,
we offer some concluding remarks on Section~\ref{sec:conclusion}.
To enhance the flow of the paper, all the proofs are collected in
the Appendices.

\section{The Three Node Wireless Network}\label{sec:general}
\begin{figure}[thb]
\centering
\includegraphics[width=0.5 \textwidth]{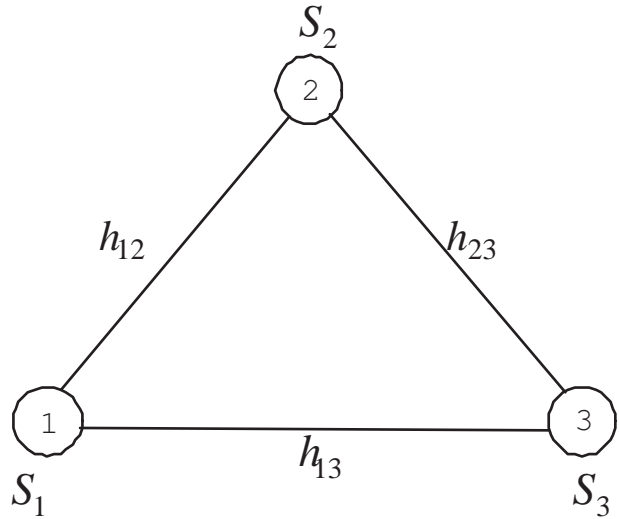}
\caption{An illustration of the three-node (half-duplex) wireless
network. Each node may be interested in a subset or all the
observation variables distributed across the network.}
\label{fig:conferencechannel}
\end{figure}

Figure~\ref{fig:conferencechannel} illustrates a network
consisting of three nodes each observing a different source. In
the general case, the three sources can be correlated. Nodes are
interested in obtaining a subset or all the source variables at
the other nodes. To achieve this goal, nodes are allowed to
coordinate and exchange information over the wireless channel.
Mathematically, the three node wireless network studied in this
paper consists of following elements:
\begin{enumerate}
\item The three sources $S_{i},i=1,2,3$, drawn i.i.d.\ from
certain known joint distribution $p(s_{1},s_{2},s_{3})$ over a
{\bf finite} set
$\mathcal{S}_{1}\times\mathcal{S}_{2}\times\mathcal{S}_{3}$. We
denote by $S_i^K$ the length-$K$ discrete source sequence
$S_i(1),\ldots,S_i(K)$ at the $i$-th node. Throughout the sequel,
we use capital letters to refer to random variables and small
letters for realizations.

\item We consider the discrete-time {\em additive white Gaussian
noise} (AWGN) channel. At time instant $n$, node $j$ receives
\begin{equation}
Y_{j}(n)=\sum \limits_{i\neq j}h_{ji}X_{i}(n)+Z_{j}(n)
\end{equation}
where $X_{i}(n)$ is the transmitted signal by node-$i$ and
$h_{ji}$ is the channel coefficient from node $i$ to $j$. To
simplify the discussion, we assume the channel coefficients are
symmetric, i.e., $h_{ij}=h_{ji}$. These channel gains are assumed
to be known {\em a-priori} at the three nodes. We also assume that
the additive zero-mean Gaussian noise is spatially and temporally
white and has the same unit variance ($\sigma^2=1$).

\item We consider half-duplex nodes that cannot transmit and
receive {\em simultaneously} using the same degree of freedom.
Without loss of generality, we split the degrees of freedom
available to each node in the temporal domain, so that, at each
time instant $n$, a node-$i$ can either transmit ({\em T-mode},
$Y_i(n)=0$) or receive ({\em R-mode}, $X_i(n)=0$), but never both.
Due to the half-duplex constraint, at any time instant, the
network nodes are divided into two groups: the T-mode nodes
(denoted by $\mathcal{T}$) and the R-mode nodes ($\mathcal{R}$). A
partition $(\mathcal{T},\mathcal{R})$ is called a network state.

\item Let $P_i^{(l)}$ denote the average transmit power at the
$i$-th node during the $m_l$ network state. We adopt a short-term
power constraint such that the total power of all the T-mode nodes
at any network state is limited to $P$, that is,
\begin{equation}\label{equ:gpowercon}
\sum\limits_{i\in\mathcal{T}_l }{P_{i}^{(l)}}\leq P, \quad \forall
m_l.
\end{equation}

\item We associate with node-$i$ an index set $I_{i}$, such that
$j\in I_{i}$ indicates that node-$i$ is interested in obtaining
$S_j$ from node-$j$ ($j\neq i$).

\item At node-$i$, a causal joint source-channel encoder converts
a length-$K$ block of source sequence into a length-$N$ codeword.
The encoder output at time $n$ is allowed to depend on the
received signal in the previous $n-1$ instants, i.e.,
\begin{equation}
X_{i}(n)=f_{i}(n,S_{i}^K,Y_{i}^{n-1}).
\end{equation}
In the special case of a separate source-channel coding approach,
the encoder decomposes into:
\begin{itemize}
\item A source encoder $f_{si}$ maps $S_i^K$ into a node message
$W_i$, i.e., $W_{i}=f_{si}(S_{i}^K)$, $W_{i}\in [1,M_{i}]$. \item
A channel encoder $f_{ci}(n)$ encodes the node message into a
channel input sequence $X_{i}(n)=f_{ci}(n,W_{i},Y_{i}^{n-1})$.
\end{itemize}

\item At node-$i$, decoder $d_i$ estimates the source variables
indexed by $I_i$
\begin{equation}
\{\hat{S}_{ij}^K\}=d_{i}(Y_{i}^N,S_{i}^K), \quad \forall j\in
I_{i}
\end{equation}
where $\hat{S}_{ij}^K$ denotes the estimation of $S_j^K$ at node
$i$. In the case of a separate source-channel coding scheme,
decoder $d_{i}$ consists of the following:
\begin{itemize}
\item A channel decoder $d_{ci}$, $\hat{W}_{ij}=d_{ci}(Y_{i}^N)$.
\item A source decoder $d_{si}$,
$\hat{S}_{ij}^K=d_{si}(\hat{W}_{ij},S_{i}^K)$.
\end{itemize}

\item  A decoding error is declared if any node fails to
reconstruct its intended source variables correctly. Thus, the
{\bf joint} error probability can be expressed as
\begin{equation} P_{e}^{N,K}=\text{Prob}\{\bigcup\limits_{j\in
I_{i},i=1,2,3}\{\hat{S}_{ij}^K\neq S_{j}^K\}\}.
\end{equation}
In the case of a separate coding scheme, the error probability
reduces to
\begin{equation}
P_{e}^{N}=\text{Prob}\{\bigcup\limits_{j\in
I_{i},i=1,2,3}\{\hat{W}_{ij}\neq W_{j}\}\}.
\end{equation}

\item An efficient cooperation strategy should strive to maximize
the achievable rate given by $\frac{KH(S_1,S_2,S_3)}{N}$, where
$N$ is the minimum number of channel uses necessary to satisfy the
network requirements. For a fixed $H(S_1,S_2,S_3)$, this
optimization is equivalent to minimizing the bandwidth expansion
factor $\tau=\frac{N}{K}$\footnote{The bandwidth expansion factor
terminology is motivated by the real time application where the
bandwidth of the channel must be $N/K$ times the bandwidth of the
source process.}. Due to a certain additive property, using the
bandwidth expansion factor will be more convenient in the
conference channel scenario. A bandwidth expansion factor $\tau$
is said to be achievable if there exists a series of
source-channel codes with $N,K\rightarrow\infty$ but
$\frac{N}{K}\rightarrow\tau$, such that $P_{e}^{N,K}\rightarrow
0$. In the feedback-relay and multicast channel, minimizing the
bandwidth expansion factor reduces to the more conventional
concept of maximizing the rate given by $R=\frac{\log_2(M)}{N}$,
where $M$ is the size of message set at the source node. \item
Throughout the sequel we will use the shorthand notation
\begin{equation}
C(x)=\frac{1}{2}\log\left(1+x\right).
\end{equation}
\end{enumerate}

The three-node network model encompasses many important network
communication scenarios with a wide range of complexity,
controlled by various configurations of the index sets and the
sources. From this perspective, the relay channel represents the
simplest situation where one node serves as the relay for the
other source-destination pair, e.g.,
$\mathcal{S}_2=\mathcal{S}_3=\phi$, $I_1=I_2=\phi$ and
$I_3=\{1\}$. If we enlarge the index set $I_2=\{1\}$, meaning
node-$2$ now is also interested in obtaining the source message,
then the problem becomes the multicast channel. Furthermore, if
the two receivers (node-$2$ and $3$) in the multicast case have
additional observations, i.e., $S_2$ and $S_3$, which are
correlated with the source variable $S_1$, then the problem
generalizes to the so-called multicast with side information
scenario. We refer to the most complex scenario as the conference
channel. In this scenario, the three sources are correlated and
every node attempts to reconstruct the other two sources, i.e.,
$I_i=\{1,2,3\}-\{i\}$. While it is easy to envision other variants
of the three node network, we decide to limit ourselves to these
special cases. This choice stems from our belief that other
scenarios do not add further insights to our framework. For
example, another variant of the feedback-relay channel would allow
the relay to observe its own side information. Careful
consideration of this case, however, shows that our analysis in
Section~\ref{sec:relay} extends to this case with only slight
modifications. Similarly, inspired by our modular approach for the
conference channel, one can decompose the multiple-access channel
with correlated sources into two stages of feedback-relay channels
with side information.

\section{The Feedback-Relay Channel}
\label{sec:relay}

Our formulation for the three node network allows for a more
realistic investigation of the relay channel with feedback. In
this scenario, node-$1$ is designated as the source node, node-$3$
the destination, and node-$2$ the relay. Since there is only one
source in this case, one can easily see that maximizing the
achievable rate $R$ from source to destination is equivalent to
minimizing the bandwidth expansion factor. Before proceeding to
our scenario of interest, we review briefly the available results
on the AWGN relay channel.

In a recent work \cite{Kramer.relay}, Kramer {\em et al.} present
a comprehensive overview of existing cooperation strategies, and
the corresponding achievable rates, for full-duplex/half-duplex
relay channels. In our work, we focus on two classes of
cooperation strategies, namely 1) Decode and Forward (DF) and 2)
Compress and Forward (CF) strategies.\footnote{For simplicity of
presentation, in our notation, we do not distinguish between
``partial'' DF and ``complete'' DF strategies (The same applies to
CF). For more details the reader is referred to~\cite{T.Cover}.}

In DF cooperation, the relay node first decodes the source message
and then starts aiding the destination node in decoding (through a
beamforming approach). More specifically, the transmission cycle
is divided into two stages. In the first stage, which occupies a
fraction $t$ of the total time, the source node sends common
messages to both the relay and destination node. Typically more
information is sent in this stage than can the destination node
decode. Having successfully decoded the source message in this
stage, the relay node uses the second stage to help the
destination resolve its uncertainty about the transmitted
codeword. During the second stage, a new message is also sent to
the destination node from the source node, along with the
information from the relay. When the source-relay link is very
noisy, one can argue that requiring the relay node to decode the
message before starting to help the destination may, in fact,
adversely affect performance. The CF strategy avoids this drawback
by asking the relay to ``compress'' its observations and send it
to the destination. In this approach, Wyner-Ziv source compression
is employed by the relay to allow the destination node to obtain a
(noisy) copy of the relay observations. Similar to the DF
strategy, the transmission cycle is divided into two stages.
During the first stage, both the relay and the destination listen
to the source node. The relay then quantizes its observations and
sends the quantized data to the destination node during the second
stage. In general, the correlation between the relay observations
and the destination observations can be exploited by the Wyner-Ziv
coding to reduce the data rate at the relay node. During the
second stage, new information is also sent by the source that
further boosts the total throughput. Here we omit the detailed
proofs and refer the interested readers to the relevant works
\cite{T.Cover,Gamal.relay,Kramer.relay,Moha.cheaprelay,
Moha.gaucheaprelay, Madsen.relay,Kram.half,Madsen.compress}. We
note, however, that the statement of the results allows for
employing optimal power allocation policies to maximize the
throughput.

\begin{lem}\label{lem:df-cf}
The achievable rate of the DF and CF strategies are given by
\begin{equation} \label{eq:R_df}
\begin{split}
R_{DF}=\sup_{t,r_{12},P_i^{(j)}}\min\biggl\{&
tC\Bigl(h_{12}^2P\Bigr)+
(1-t)C\Bigl((1-r_{12}^2)h_{13}^2P_1^{(2)}\Bigr);\\
&tC\Bigl(h_{13}^2P\Bigr) +(1-t)C\Bigl(h_{13}^2P_1^{(2)}+
2r_{12}h_{13}h_{23}\sqrt{P_1^{(2)}P_2^{(2)}}+h_{23}^2P_2^{(2)}\Bigr)
\biggr\}.
\end{split}
\end{equation}

\begin{equation} \label{eq:R_cf}
R_{CF}=\sup_{t,P_i^{(j)}}
tC\Bigl(\bigl(h_{13}^2+\frac{h_{12}^2}{1+\sigma_2^2} \bigr)P\Bigr)
+ (1-t)C\Bigl(h_{13}^2P_1^{(2)}\Bigr).
\end{equation}
where
\begin{equation}
\sigma_2^2=\frac{(h_{12}^2+h_{13}^2)P+1}{(h_{13}^2P+1)
\biggl(\Bigl(1+\frac{h_{23}^2P_2^{(2)}}{h_{13}^2P_1^{(2)}+1}\Bigr)^{
\frac{1-t}{t}}-1\biggr)},
\end{equation}
and
\begin{equation}
P_{1}^{(2)}+P_{2}^{(2)}=P.
\end{equation}
\end{lem}

\begin{figure}[thb]
\centering
\includegraphics[width=0.7 \textwidth]{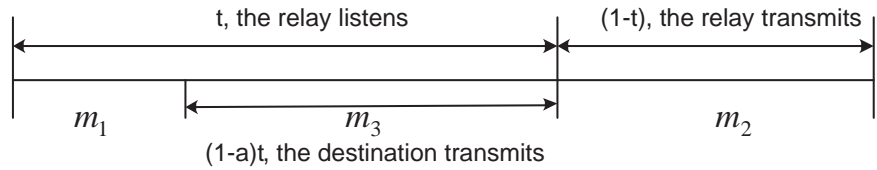}
\caption{The operation sequence of the half-duplex relay channel
with realistic feedback.} \label{fig:relaystate}
\end{figure}

We are now ready to present the cooperation strategy for the relay
channel with feedback. In a nutshell, the proposed strategy
combines the DF and CF strategies to overcome the bottleneck of a
noisy source-relay channel. In this FeedBack (FB) approach, the
destination first assists the relay in decoding via CF
cooperation.  After decoding, the relay starts helping the
destination via a DF configuration. Due to the half-duplex
constraint, every cycle of transmission is divided into the
following three stages (as shown in Fig.~\ref{fig:relaystate}).

\begin{itemize}
\item The first state lasts for a fraction $\alpha t$ of the cycle
($0\leq t,\alpha\leq 1$). In this stage, both the relay and the
destination listen to the source. We refer to the network state in
this stage as $m_1$. \item The feedback stage lasts for a fraction
$(1-\alpha) t$ of the cycle. In this stage, the relay listens to
both the destination and the source. Since the destination is not
yet able to completely decode the source message, it sends to the
relay node a Wyner-Ziv compressed version of its observations. We
refer to the network state in this stage as $m_3$. \item The final
stage lasts for a fraction $(1-t)$ of the cycle. Having obtained
source information, the relay is now able to help the destination
node in decoding the source message. We refer to the network state
in this stage as $m_2$.
\end{itemize}

The time-division parameters $t$ and $\alpha$ control the relative
duration of each network state. In particular, $t$ represents the
total time when the relay node is in the receive mode. The
feedback parameter $\alpha$ controls the amount of feedback, i.e.,
a $(1-\alpha)$ fraction of the total relay listening time is
dedicated to feedback. Here, we stress that this formulation for a
relay channel with feedback represents a ``realistic'' view that
attempts to capture the constraints imposed by the wireless
scenario (as opposed to the {\em ideal} feedback assumed in
existing works, e.g., \cite{T.Cover}). The feedback considered
here simply refers to transmission from the destination to relay
over the same (noisy) wireless channel. Using random coding
arguments we obtain the following achievable rate for the proposed
feedback scheme.

\begin{lem}\label{lem:fb}
The achievable rate of the feedback scheme is given by
\begin{equation} \label{eq:R_fb}
\begin{split}
R_{FB}=\sup_{\alpha,t,r_{12},P_i^{(j)}}\min\Biggl\{& \alpha
tC\Bigl(\bigl(\frac{h_{13}^2}{1+\sigma_3^2}+h_{12}^2\bigr)
P_1^{(1)}\Bigr) +(1-\alpha)tC\Bigl(h_{12}^2P_1^{(3)}\Bigr) \\&+
(1-t)C\Bigr((1-r_{12}^2)h_{13}^2P_1^{(2)}\Bigr); \\
&\alpha tC\Bigl(h_{13}^2P_1^{(1)}\Bigr) +
(1-t)C\Bigl(h_{13}^2P_1^{(2)}+2r_{12}h_{13}h_{23}\sqrt{P_1^{(2)}
P_2^{(2)}}+h_{23}^2P_2^{(2)}\Bigr) \Biggr\}
\end{split}
\end{equation}
where
\begin{equation}
\sigma_3^2=\frac{\bigl(h_{12}^2+h_{13}^2\bigr)P_1^{(1)}+1}{
\bigl(h_{12}^2P_1^{(1)}+1\bigr)\biggl(\Bigl(1+\frac{h_{23}^2P_3^{(3)}}{
h_{12}^2P_1^{(3)}+1}\Bigr)^{\frac{1-\alpha}{\alpha}}-1\biggr) }
\end{equation}
and $r_{12}$ is the correlation between $X_1,X_2$ during state
$m_{2}$. In the proposed strategy, the total power constraint
specializes to
\begin{equation}\label{equ:relaypow}
P_1^{(1)}=P,\quad P_1^{(2)}+P_2^{(2)}=P, \quad
P_1^{(3)}+P_3^{(3)}=P.
\end{equation}
\end{lem}
\begin{proof}
Please refer to Appendix~\ref{appex:fb}.
\end{proof}

Armed with Lemmas~\ref{lem:df-cf}~and~\ref{lem:fb}, we can now
contrast the performance of the DF, CF, and FB strategies. Our
emphasis is to characterize the fundamental properties of the
feedback scheme and quantify the gain offered by it under
different assumptions on the channel gains and total power. The
{\em relay-off} performance, i.e., $R_{ro}=C(h_{13}^2P)$, serves
as a lower bound on the achievable rate. In fact, the relay-off
benchmark can be viewed as a special case of the three cooperative
schemes. For example, setting $P_2^{(2)}=0$ and $t=0$ effectively
reduces both DF and CF strategies to the relay-off case.
Therefore, one can conceptually describe the order of containment
of various schemes as ``relay-off $\subset$ DF $\subset$ FB'' and
``relay-off $\subset$ CF''. As for the performance upper bounds,
the cut-set bounds \cite{T.Cover} give rise to 1) a
multi-transmitter rate $R_{(1,2)-3}=C((h_{13}^2+h_{23}^2)P)$
corresponding to perfect cooperation between the source and relay
nodes; and 2) a multi-receiver rate
$R_{1-(2,3)}=C((h_{13}^2+h_{12}^2)P)$ corresponding to perfect
cooperation between the relay and destination nodes.

\begin{figure}[thb]
\begin{center}
\includegraphics[width=0.3\textwidth]{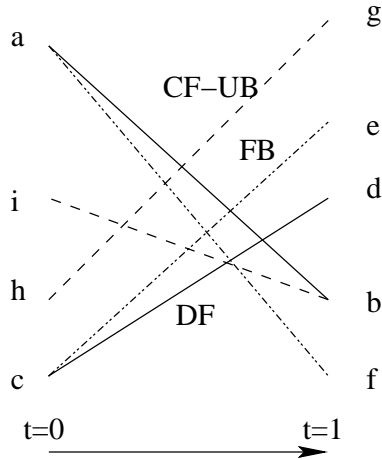}
\caption{\label{fig:linebd} A geometric representation of FB-, DF-
and CF-relay schemes. The solid lines are for $R_{DF}$ in
\eqref{eq:R_df}, the dash-dotted for $R_{FB}$ in \eqref{eq:R_fb},
and the dashed for the upperbound of $R_{CF}$ in
$\eqref{eq:R_cf_ub}$. The various endpoints in the figure are (a)
$C(h_{13}^2P_1^{(2)}+2r_{12}h_{13}h_{23}\sqrt{P_1^{(2)}P_2^{(2)}}
+h_{23}^2P_2^{(2)})$, (b) $C(h_{13}^2P_1^{(1)})$, (c)
$C((1-r_{12}^2) h_{13}^2P_1^{(2)})$, (d) $C(h_{12}^2P_1^{(1)})$,
(e) $\alpha C(( \frac{h_{13}^2}{1+\sigma_3^2}+h_{12}^2)P_1^{(1)})
+ (1-\alpha) C(h_{12}^2P_1^{(3)})$, (f) $\alpha
C(h_{13}^2P_1^{(1)})$, (g) $C((h_{13}^2+h_{12}^2)P_1^{(1)})$, (h)
$C(h_{13}^2P_1^{(2)})$, and (i)
$C(h_{13}^2P_1^{(2)}+h_{23}^2P_2^{(2)})$. }
\end{center}
\end{figure}

The achievable rate for the DF strategy, i.e., $R_{DF}$, enjoys an
intuitive geometric interpretation: each expression within the min
operator is a linear segment in the parameter $t\in[0,1]$ (see
\eqref{eq:R_df}). Hence, the optimal time $t$, assuming the other
variables remain fixed, can be simply determined by the
intersection point of the two associated line segments, as
illustrated in Fig.~\ref{fig:linebd}. On the other hand, $R_{FB}$
and $R_{CF}$ are characterized by more complicated expressions due
to the dependency of $\sigma_3^2$ and $\sigma_2^2$ upon the
time-division parameters. Our next result finds upper bounds on
$R_{FB}$ and $R_{CF}$ which allow for the same simple
line-crossing interpretation as $R_{DF}$.

\begin{lem}\label{lem:up_bds}
The achievable rate of the feedback scheme is upper bounded by
\begin{equation} \label{eq:R_fb_ub}
\begin{split}
R_{FB}\leq\sup_{\alpha,t,r_{12},P_i^{(j)}}\min \Biggl\{& \alpha
tC\Bigl(h_{12}^2 P_1^{(1)}\Bigr) +(1-\alpha)t
C\Bigl(h_{12}^2P_1^{(3)}+h_{23}^2P_3^{(3)}\Bigr) +
(1-t)C\Bigr((1-r_{12}^2)h_{13}^2P_1^{(2)}\Bigr); \\
&\alpha tC\Bigl(h_{13}^2P_1^{(1)}\Bigr) +
(1-t)C\Bigl(h_{13}^2P_1^{(2)}+2r_{12}h_{13}h_{23}\sqrt{P_1^{(2)}
P_2^{(2)}}+h_{23}^2P_2^{(2)}\Bigr) \Biggr\}.
\end{split}
\end{equation}
The achievable rate of compress-and-forward is bounded by
\begin{equation} \label{eq:R_cf_ub}
\begin{split}
R_{CF}\leq\sup_{t,P_i^{(j)}}\min \Biggl\{& tC\Bigl(h_{13}^2
P_1^{(1)}\Bigr)
+ (1-t)C\Bigl(h_{13}^2P_1^{(2)}+h_{23}^2P_2^{(2)}\Bigr); \\
& tC\Bigl(\bigl(h_{13}^2+h_{12}^2\bigr)P_1^{(1)}\Bigr)+
(1-t)C\Bigl(h_{13}^2P_1^{(2)}\Bigr) \Biggr\}.
\end{split}
\end{equation}
\end{lem}
\begin{proof}
Please refer to Appendix~\ref{appex:up_bds}.
\end{proof}

Interestingly, Fig.~\ref{fig:linebd} encodes a great deal of
information regarding the performance of the three schemes. For
example, when $h_{12}^2\leq h_{13}^2$, the intersection point
corresponding to decode-forward would fall below the flat line
$C(h_{13}^2P)$ associated with the relay-off rate. More
rigorously, we have the following statement.

\begin{thm} \label{thm:relaycomph}
\begin{enumerate}
\item If $h_{12}^2\leq h_{13}^2$ then $R_{DF}\leq R_{ro}$. \item
If $h_{23}^2\leq h_{13}^2$ then $R_{CF}\leq R_{ro}$. \item If
$h_{23}^2\leq h_{12}^2$ then $R_{FB}\leq R_{DF}$.
\end{enumerate}
\end{thm}
\begin{proof}
Please refer to Appendix~\ref{appex:relaycomph}.
\end{proof}

Theorem~\ref{thm:relaycomph} reveals the fundamental impact of
channel coefficients on the performance of the different
cooperation strategies. In particular, the DF strategy is seen to
work well with a ``strong'' source-relay link. If, at the same
time, the relay-destination link is stronger, then one may exploit
feedback, i.e., $\alpha\neq 1$, to improve performance. The next
result demonstrates the asymptotic optimality of the feedback
scheme in the limit of large $h_{12}$ or $h_{23}$.

\begin{thm}\label{thm:relayasymh}
\begin{enumerate}
\item As $h_{12}$ increases, both DF- and FB-scheme achieve the
optimal beam-forming benchmark, while CF-scheme is limited by a
sub-optimal rate $C\bigl(\max\{h_{13}^2,h_{23}^2\}P\bigr)$. \item
As $h_{23}$ increases, both CF- and FB-scheme achieve the optimal
multi-receiver benchmark, while DF-scheme only approaches to a
sub-optimal rate $C\bigl(\max\{h_{12}^2,h_{13}^2\}P\bigr)$.
\end{enumerate}
\end{thm}

The proof of Theorem~\ref{thm:relayasymh} is a straightforward
limit computation, and hence, is omitted for brevity. So far we
have kept the total power $P$ constant. But in fact, the
achievable rate as a function of $P$ offers another important
dimension to the problem. First, we investigate the low power
regime, which is greatly relevant to the wide-band scenario. In
this case we study the slope $S$ of the achievable rate with
respect to $P$ ( i.e., $R\sim\frac{1}{2}(\log e)SP$). Note that
the relay-off benchmark has a slope $h_{13}^2$.

\begin{thm} \label{thm:relaylowP}
Let
$f_1(\theta,r_{12},h_{13},h_{23})=h_{13}^2\cos^2\theta+2r_{12}h_{13}h_{23}
\cos\theta\sin\theta+h_{23}^2\sin^2\theta$ and
$f_2(\theta,r_{12},h_{13}) =(1-r_{12}^2)h_{13}^2\cos^2\theta$ be a
shorthand notation, then
\begin{enumerate}

\item When $h_{12}^2\geq h_{13}^2$
\begin{equation} \label{eq:S_df}
S_{DF}=\max_{\theta,r_{12}}\frac{f_1(\theta,r_{12},h_{13},h_{23})h_{12}^2-
f_2(\theta,r_{12},h_{13})h_{13}^2}{f_1(\theta,r_{12},h_{13},h_{23})+h_{12}^2-
f_2(\theta,r_{12},h_{13})-h_{13}^2},
\end{equation}
and
\begin{equation} \label{eq:S_df_bd}
\frac{(h_{13}^2+h_{23}^2)h_{12}^2}{h_{23}^2+h_{12}^2}\leq S_{DF}
\leq
\frac{(h_{13}^2+h_{23}^2)h_{12}^2-h_{13}^4}{h_{23}^2+h_{12}^2-h_{13}^2}.
\end{equation}
\item $S_{CF} = h_{13}^2$ with $t_{opt}\rightarrow 1$.

\item $S_{FB}=S_{DF}$ with $\alpha_{opt}\rightarrow 1$.
\end{enumerate}
\end{thm}
\begin{proof}
Please refer to Appendix~\ref{appex:relaylowP}.
\end{proof}

It follows from Theorem~\ref{thm:relaylowP} that given
$h_{12}^2\geq h_{13}^2$, DF cooperation delivers a larger slope
than the relay-off, i.e.,
\begin{equation}
S_{DF}-h_{13}^2\geq\frac{h_{23}^2(h_{12}^2-h_{13}^2)}{h_{23}^2+h_{12}^2}
\geq 0.
\end{equation}
However, CF cooperation does not yield any gain in the low power
regime. Similarly, we see that the CF stage of the proposed FB
becomes useless, and hence, the scheme reduces to the DF approach
in the low power regime. The reason lies in the fact that for
small $P$, the channel output is dominated by the noise, and
hence, the compression algorithm inevitably operates on the noise,
resulting in diminishing gains.

We next quantify the SNR gain of the three schemes in the high
power regime, that is, to characterize $R\sim \frac{1}{2}\log P+
\frac{1}{2} G$ as $P\rightarrow\infty$.

\begin{thm} \label{thm:relayhighP}
Following the same shorthand notations as in
Theorem~\ref{thm:relaylowP}, we obtain
\begin{enumerate}
\item Given $h_{12}^2\geq h_{13}^2$,
\begin{equation}
G_{DF}=\max_{\theta,r_{12}}\frac{\log
f_1(\theta,r_{12},h_{13},h_{23}) \cdot\log h_{12}^2- \log
f_2(\theta,r_{12},h_{13})\cdot\log h_{13}^2}{\log
\bigl[f_1(\theta,r_{12},h_{13},h_{23}) h_{12}^2\bigr] -
\log\bigl[f_2(\theta,r_{12},h_{13})h_{13}^2\bigr]}.
\end{equation}
\item
\begin{equation}
G_{CF}=\max_{t,\theta} t\log\Bigl(h_{13}^2+\frac{h_{12}^2}{1+
\sigma_2^2(\infty)}\Bigr) +
(1-t)\log\bigl(h_{13}^2\cos^2\theta\bigr)
\end{equation}
where
\begin{equation}
\sigma_2^2(\infty)=\frac{h_{12}^2+h_{13}^2}{h_{13}^2\Bigl(\bigl(1+
\frac{h_{23}^2}{h_{13}^2}\tan^2\theta\bigr)^{\frac{1-t}{t}}-1\Bigr)}.
\end{equation}
\item $G_{FB} = G_{DF}$ with $\alpha_{opt}\rightarrow 1$.
\end{enumerate}
\end{thm}
\begin{proof}
Please refer to Appendix~\ref{appex:relayhighP}.
\end{proof}

Theorem~\ref{thm:relayhighP} reveals the fact that strict feedback
($\alpha\neq 1$) does not yield a gain in high power regime. The
reason for this behavior can be traced back to the half-duplex
constraint. When $\alpha\neq 1$, the destination spends a fraction
$(1-\alpha)t$ of time transmitting to the relay, which cuts off
the time in which it would have been listening to the source in
non-feedback schemes. Such a time loss reduces the pre-log
constant, which cannot be compensated by the cooperative gain when
P becomes large.

\begin{figure}[htb]
\begin{center}
\includegraphics[width=0.4 \textwidth]{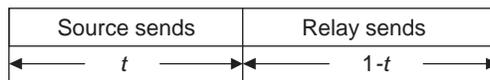}\caption{
Orthogonal transmission: the source spends $t$ part of the time to
transmit, the relay uses the remaining $1-t$ part of the time to
transmit.}\label{fig:orthogonal}
\end{center}
\end{figure}

At this point, we wish to make a side comment contrasting the
half-duplex constraint with orthogonal relay channels. The
orthogonal cooperation framework was recently proposed as a
practical way to address the half-duplex requirement
\cite{Gamal.relay,Laneman.halfduplex}. For simplicity, let's
consider the non-feedback scenario and assume that the available
bandwidth is $W$ Hz, and hence, the total resources available to
every node in network is $2W$ real dimensions per second. The
half-duplex constraint dictates {\bf only} orthogonality at each
node, where the available degrees of freedom, in the
time-frequency plane, is splitted into two parts. The node uses
the first part to receive and the second to transmit. In the
orthogonal cooperation approach, however, one imposes
orthogonality at the network level (i.e., no two nodes can now
transmit in the same degree of freedom). In particular, the
channel is split into two sub channels (either in time
domain~\cite{Laneman.halfduplex} or frequency
domain~\cite{Gamal.relay}), where the source uses one of the
sub-channels to transmit information to the relay and destination,
and the relay uses the other sub-channel to transmit to the
destination. One can now see that this orthogonalization is
sufficient but {\bf not} necessary to satisfy the half-duplex
constraint. Figure~\ref{fig:orthogonal} shows the orthogonal
cooperation scheme which splits the channel in the time domain.
When relay sends, it can use either the DF or CF strategies. Using
the same argument as in the previous part, one obtains the
following achievable rate for the orthogonal DF strategy scheme,
\begin{equation}
R_{ODF}=\max\limits_{t}\min\{tC(h_{12}^2P);\;tC(h_{13}^2P)+(1-t)C(h_{23}^2P)\}.
\end{equation}

It is now clear that $R_{ODF}$ is just a special case of $R_{DF}$
where, for any $t$, one can obtain the corresponding $R_{ODF}$ by
setting $P_{1}^{(2)}=0$ in~\eqref{eq:R_df}. One can use the fact
that $P_{1}^{(2)}=0$ is not necessarily the optimal power
assignment that maximizes~\eqref{eq:R_df} to argue for the
sub-optimality of the orthogonal DF strategy (i.e., $R_{DF}\geq
R_{ODF}$). More generally, the same argument can be used to
establish the sub-optimality of any orthogonal cooperation
strategy.

\begin{figure}
\centering
\includegraphics[width=0.7 \textwidth]{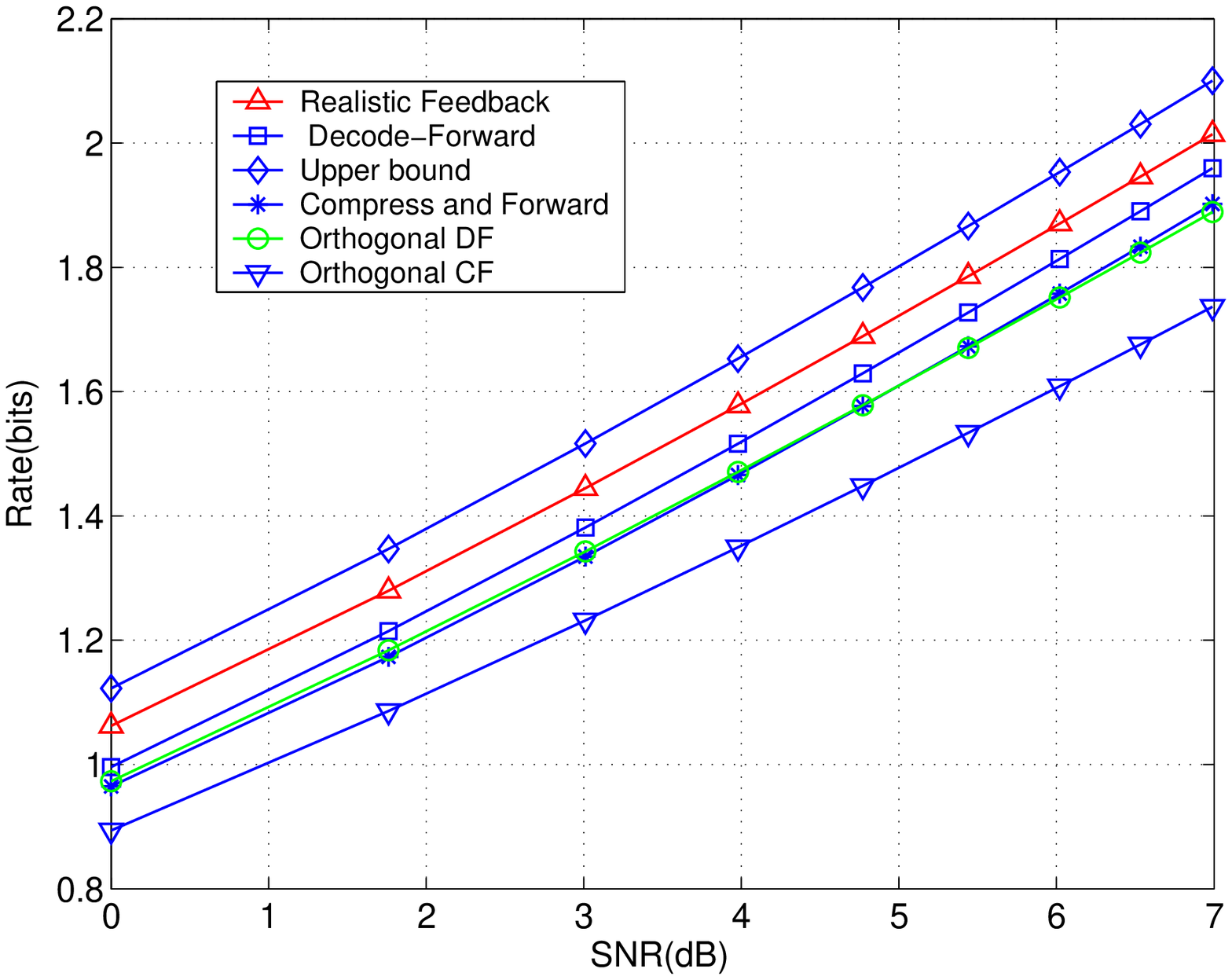}
\caption{The achievable rate of various schemes in the half-duplex
relay channel, $h_{12}=1.8,h_{13}=1,h_{23}=23 dB$.
}\label{fig:ratevsp}
\end{figure}

\begin{figure}
\centering
\includegraphics[width=0.7\textwidth]{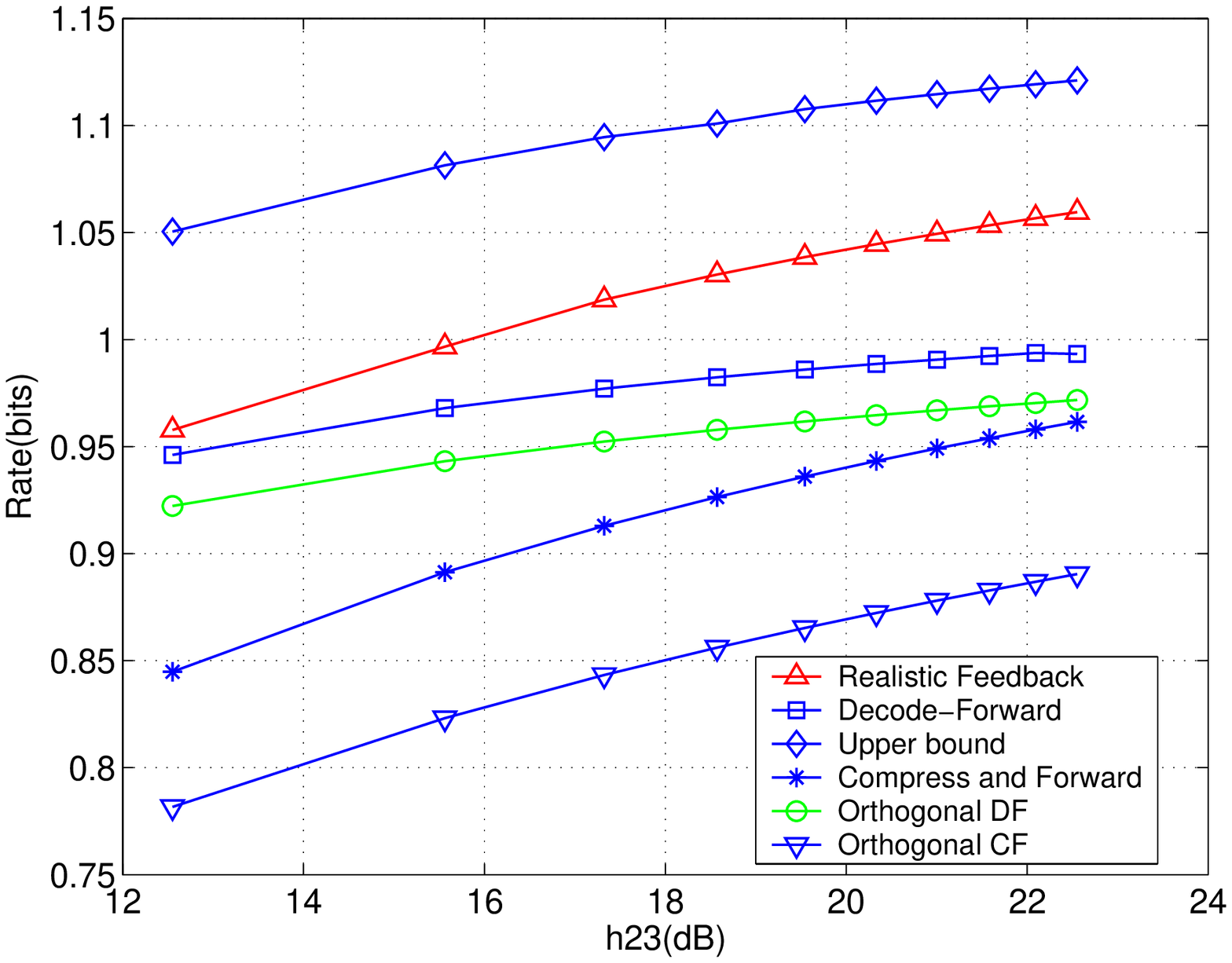}
\caption{The achievable rate of various schemes in the half-duplex
relay channel, $h_{12}=1.8,h_{13}=1,\text{SNR}=0dB$.
}\label{fig:ratevsh23}
\end{figure}

\begin{figure}[htb]
\begin{center}
\includegraphics[width=0.7 \textwidth]{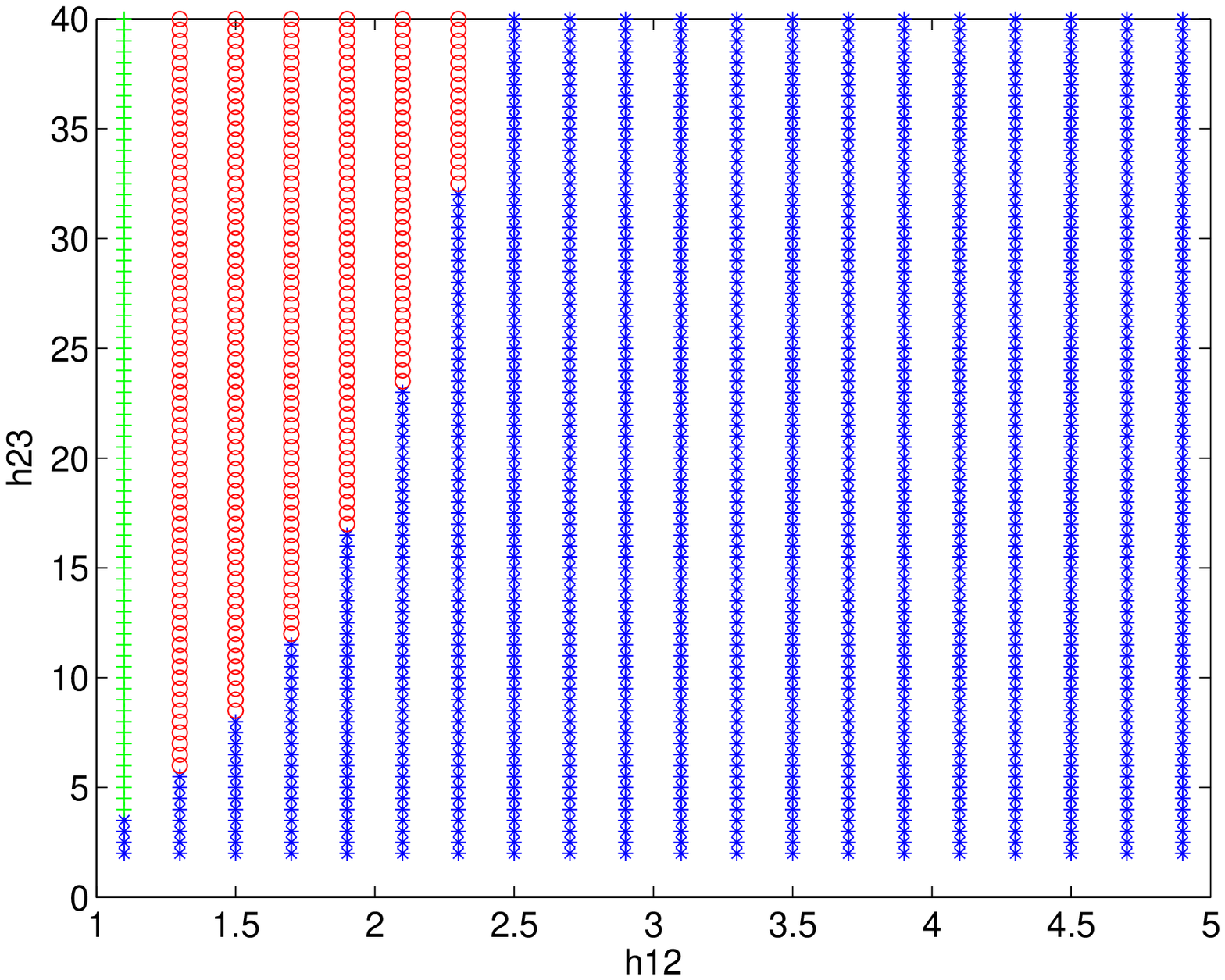}
\caption{\label{fig:relay_p2} Optimal regions for
decode-and-forward (denoted by `*'), compress-and-forward (denoted
by `+'), and feedback strategy (denoted by `o'), when
$\text{SNR}=3 dB$.}
\end{center}
\end{figure}

\begin{figure}[htb]
\begin{center}
\includegraphics[width=0.7 \textwidth]{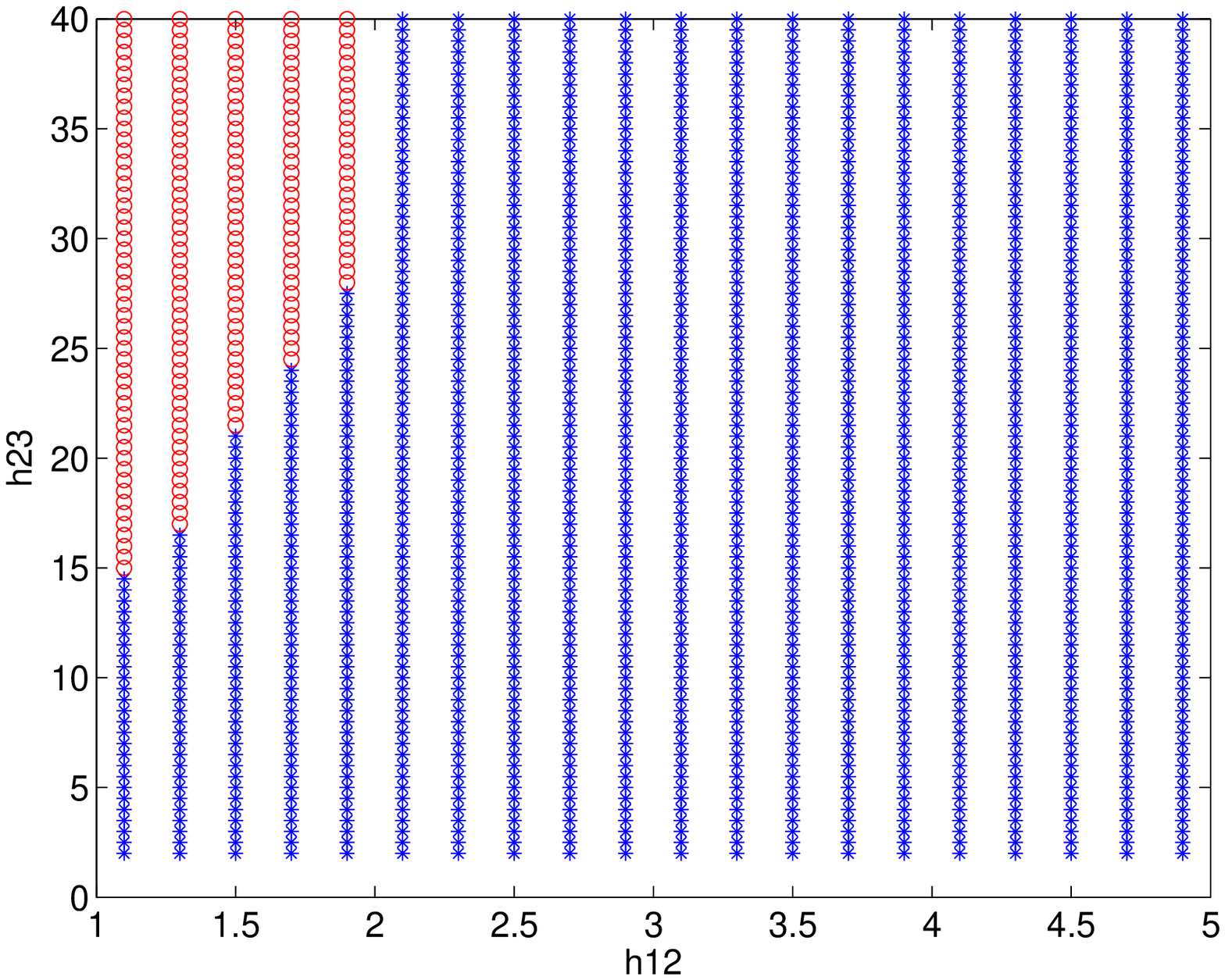}
\caption{\label{fig:relay_p001} Optimal regions for
decode-and-forward (denoted by `*'), compress-and-forward (denoted
by `+'), and feedback strategy (denoted by `o'), when
$\text{SNR}=-20 dB$.}
\end{center}
\end{figure}

\begin{figure}[htb]
\begin{center}
\includegraphics[width=0.7\textwidth]{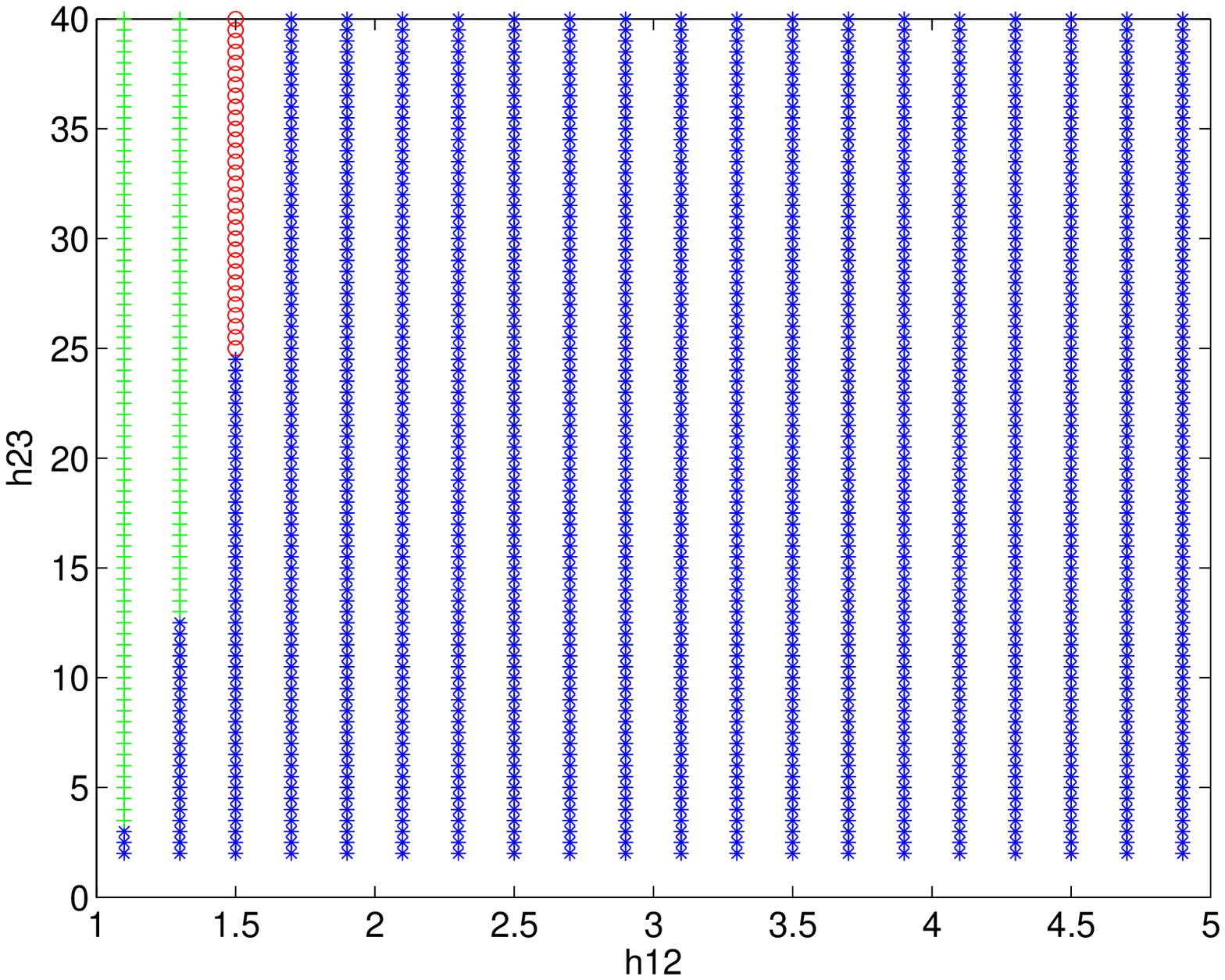}
\caption{\label{fig:relay_p100} Optimal regions for
decode-and-forward (denoted by `*'), compress-and-forward (denoted
by `+'), and feedback strategy (denoted by `o'), when
$\text{SNR}=20dB$.}
\end{center}
\end{figure}

We conclude this section with simulation results that validate our
theoretical analysis. Figure~\ref{fig:ratevsp} reports the
achievable rate of various schemes, when $h_{12}=1.8$, $h_{13}=1$,
and $h_{23}=200$. This corresponds to the case when the
source-relay channel is a little better than the
source-destination channel, and the relay-destination channel is
quite good. This is the typical scenario when feedback results in
a significant gain, as demonstrated in the figure. In the figure,
we also see the sub-optimality of orthogonal cooperation
strategies. Figure~\ref{fig:ratevsh23} reports the achievable
rates of various schemes, when $h_{12}=1.8$, $h_{13}=1$, and
$P=1$, as we vary the relay-destination channel gain $h_{23}$. We
can see that as the relay-destination channel becomes better, the
advantage of feedback increases.
Figures~\ref{fig:relay_p2},~\ref{fig:relay_p001},~\ref{fig:relay_p100}
illustrate regions in the $h_{12}-h_{23}$ plane ($h_{13}=1$)
corresponding to the best of the three strategies. It is seen that
feedback can improve upon both DF and CF strategies in certain
operating regions. However, as predicted by our analysis, such
gain diminishes when either $P\rightarrow 0$ or
$P\rightarrow\infty$. Overall, we can see that the proposed FB
cooperation scheme combines the benefits of both the DF and CF
cooperation strategies, and hence, attains the union of the
``nice'' properties of the two strategies. On the other hand, the
gain offered by feedback seems to be limited to certain operating
regions, as defined by the channel gains, and diminishes in either
the low or high power regime.

\section{The Multicast Channel}
\label{sec:multicast} The relay channel, considered in the
previous section, represents the simplest example of a three-node
wireless network. A more sophisticated example can be obtained by
requiring node-$2$ to decode the message generated at node-$1$.
This corresponds to the multicast scenario. Similar to the relay
scenario, we focus on maximizing the achievable rate from node-$1$
to both node-$2$ and $3$, without any loss of generality. The
half-duplex and total power constraints, adopted here, introduce
an interesting design challenge. To illustrate the idea, suppose
node-$2$ decides to help node-$3$ in decoding. In this case, not
only does node-$2$ compete with the source node for transmit
power, but it also sacrifices its listening time for the sake of
helping node-$3$. It is, therefore, not clear {\em a-priori} if
the network would benefit from this cooperation. In the following,
we answer this question in the affirmative and further propose a
greedy cooperation strategy that enjoys several {\em nice}
properties.

In a recent work \cite{Servetto.broadcast}, the authors considered
another variant of the multicast channel and established the
benefits of receiver cooperation in this setup. The fundamental
difference between the two scenarios is that, in
\cite{Servetto.broadcast}, the authors assumed the existence of a
dedicated link between the two receivers. This dedicated link was
used by the {\em strong} receiver to help the {\em weak} receiver
in decoding through a DF strategy. As expected, such a cooperation
strategy was shown to strictly enlarge the achievable rate region
\cite{Servetto.broadcast}. In our work, we consider a more
representative model of the wireless network in which all
communications take place over the same channel, subject to the
half-duplex and total power constraints. Despite these
constraining assumptions, we still demonstrate the significant
gains offered by receiver cooperation. Inspired by the
feedback-relay channel, we further construct a greedy cooperation
strategy that significantly outperforms the DF scheme
\cite{Servetto.broadcast} in many relevant scenarios.

In the non-cooperative scenario, both node-$2$ and node-$3$ will
listen all the time, and hence, the achievable rate is given by
\begin{equation}
C_{non-coop}=C(\min\{h_{12}^{2},h_{13}^2\}P).
\end{equation}
Due to the half-duplex constraint, time is valuable to both nodes,
which makes them selfish and unwilling to help each other. Careful
consideration, however, reveals that such a {\em greedy} approach
will lead the nodes to cooperate. The enabling observation stems
from the feedback strategy proposed for the relay channel in which
the destination was found to get a higher achievable rate if it
sacrifices some of its receiving time to help the relay. Motivated
by this observation, our strategy decomposes into three stages,
without loss of generality we assume $h_{12}^2>h_{13}^2$, 1) $m_1$
lasting for a fraction $\alpha t$ of the frame during which both
receivers listen to node-$1$; 2) $m_3$ occupying $(1-\alpha)t$
fraction of the frame during which node-$3$ sends its compressed
signal to node-$2$; and 3) $m_2$ (the rest $1-t$ fraction) during
which node-$1$ and $2$ help node-$3$ finish decoding. One major
difference between the multicast and relay scenarios is that in
the second stage the source cannot send additional (new)
information to node-$3$, for it would not be decoded by node-$2$,
thus violating the multicast requirement that both receivers
obtain the same source information. Here, we observe that the last
stage of cooperation, in which node-$2$ is helping node-$3$, is
still motivated by the greedy approach. The idea is that node-$1$
will continue transmitting the same codeword until both receivers
can successfully decode. It is, therefore, beneficial for node-$2$
to help node-$3$ in decoding faster to allow the source to move on
to the next packet in the queue. A slight modification of the
proof of Lemma~\ref{lem:fb} results in the following.

\begin{lem}\label{lem:multicast}
The achievable rate of the greedy strategy based multicast scheme
is given by
\begin{equation}
\begin{split}
R_{g}=\sup_{\alpha,t,P_i^{(j)}}\min\Biggl\{& \alpha
tC\Bigl(\bigl(\frac{h_{13}^2}{1+\sigma_4^2}+h_{12}^2\bigr)
P\Bigr) +(1-\alpha)tC\Bigl(h_{12}^2 P_1^{(3)}\Bigr); \\
&\alpha tC\Bigl(h_{13}^2P\Bigr) +
(1-t)C\Bigl((h_{13}^2+h_{23}^2)P\Bigr) \Biggr\}
\end{split}
\end{equation}
where
\begin{equation}
\sigma_4^2=\frac{\bigl(h_{12}^2+h_{13}^2\bigr)P+1}{
\bigl(h_{12}^2P+1\bigr)\biggl(\Bigl(1+\frac{h_{23}^2P_3^{(3)}}{
h_{12}^2P_1^{(3)}+1}\Bigr)^{\frac{1-\alpha}{\alpha}}-1\biggr) },
\end{equation}
and $r_{12}$ denotes the correlation between $X_{1},X_{2}$ during
state $m_{2}$. The ``sup'' operator is taken over the total power
constraint.
\end{lem}

We observe that the DF multicast scheme corresponds to the special
case of $\alpha=1$, which has a rate
\begin{equation}\label{equ:multicastgau}
R_{DF}=\sup_{t}\min\Bigl\{ tC\bigl(h_{12}^2P\bigr);\;
tC\bigl(h_{13}^2P\bigr) + (1-t)C\Bigl((h_{13}^2+h_{23}^2)P\Bigr)
\Bigr\}.
\end{equation}

The cut-set upperbounds give rise to the two following benchmarks:
beam-forming $R_{(1,2)-3}=C\bigl((h_{13}^2+h_{23}^2)P\bigr)$ and
multi-receiver $R_{1-(2,3)}=C\bigl((h_{13}^2+h_{12}^2)P\bigr)$.
Similar to the relay channel scenario, we examine in the following
the asymptotic behavior of the greedy strategy as a function of
the channel coefficients and available power.

\begin{thm} \label{Them:multicastlim}
\begin{enumerate}
\item The greedy cooperative multicast scheme strictly increases
the multicast achievable rate (as compared to the non-cooperative
scenario).

\item The greedy strategy approaches the beam-forming benchmark as
$h_{12}$ increases, i.e.,
\begin{equation}
\lim_{h_{12}\rightarrow \infty}R_{g}=C((h_{13}^2+h_{23}^2)P).
\end{equation}
\item The greedy strategy approaches the multi-receiver benchmark
as $h_{23}$ increases, i.e.,
\begin{equation}
\lim_{h_{23}\rightarrow\infty}R_{g}=C((h_{12}^{2}+h_{13}^{2})P).
\end{equation}
\item As $P\rightarrow 0$, the slope of the greedy strategy
achievable rate is given by
\begin{equation}
S_{g}=\frac{h_{12}^2(h_{23}^2+h_{13}^2)}{h_{12}^2+h_{23}^2}.
\end{equation}
\item As $P\rightarrow\infty$, the SNR gain
$G_{g}=G_{non-coop}=\log h_{13}^2$ with $t_{opt}\rightarrow 1$.
\end{enumerate}
\end{thm}
\begin{proof}
Please refer to Appendix~\ref{appex:multicast}.
\end{proof}

Parts 2), 3) demonstrate the asymptotic optimality of the greedy
multicast as the channel gains increase (the proof follows the
same line as that of Theorem~\ref{thm:relayasymh}). On the other
hand, we see that the large-power asymptotic of the multicast
channel differs significantly  from that of the relay channel. In
the relay case (Theorem~\ref{thm:relayhighP}), the contribution of
feedback diminishes ($G_{FB}=G_{DF}$) in this asymptotic scenario,
but cooperation was found to be still beneficial, that is
$G_{DF}>\log h_{13}^2$. To the contrast, the gain of receiver
cooperation in the multicast channel disappears as $P$ increases.
This is because, unlike the relay scenario, at least one receiver
must cut its listening time in any cooperative multicast scheme
due to the half-duplex constraint. Such a reduction induces a
pre-log penalty in the rate, which results in substantial loss
that cannot be compensated by cooperation as $P\rightarrow
\infty$, and hence, the greedy strategy reduces to the
non-cooperative mode automatically.

\begin{figure}[thb]
\begin{center}
\includegraphics[width=0.7\textwidth]{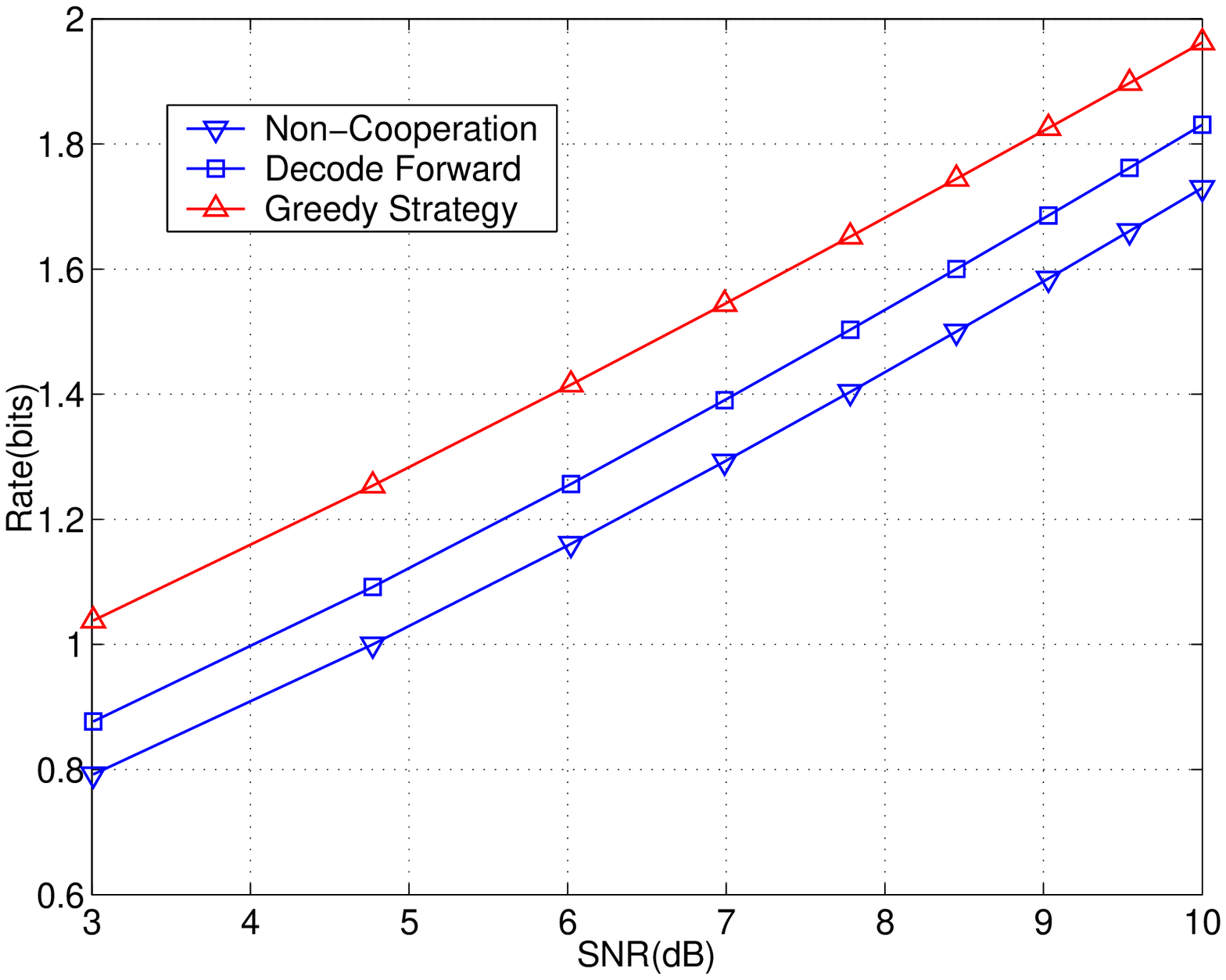}
\caption{The achievable rate of various schemes in the multicast
channel, $h_{12}=1.1$, $h_{13}=1$, and $h_{23}=23dB$. }
\label{fig:multicastpchange}
\end{center}
\end{figure}

\begin{figure}[thb]
\begin{center}
\includegraphics[width=0.7 \textwidth]{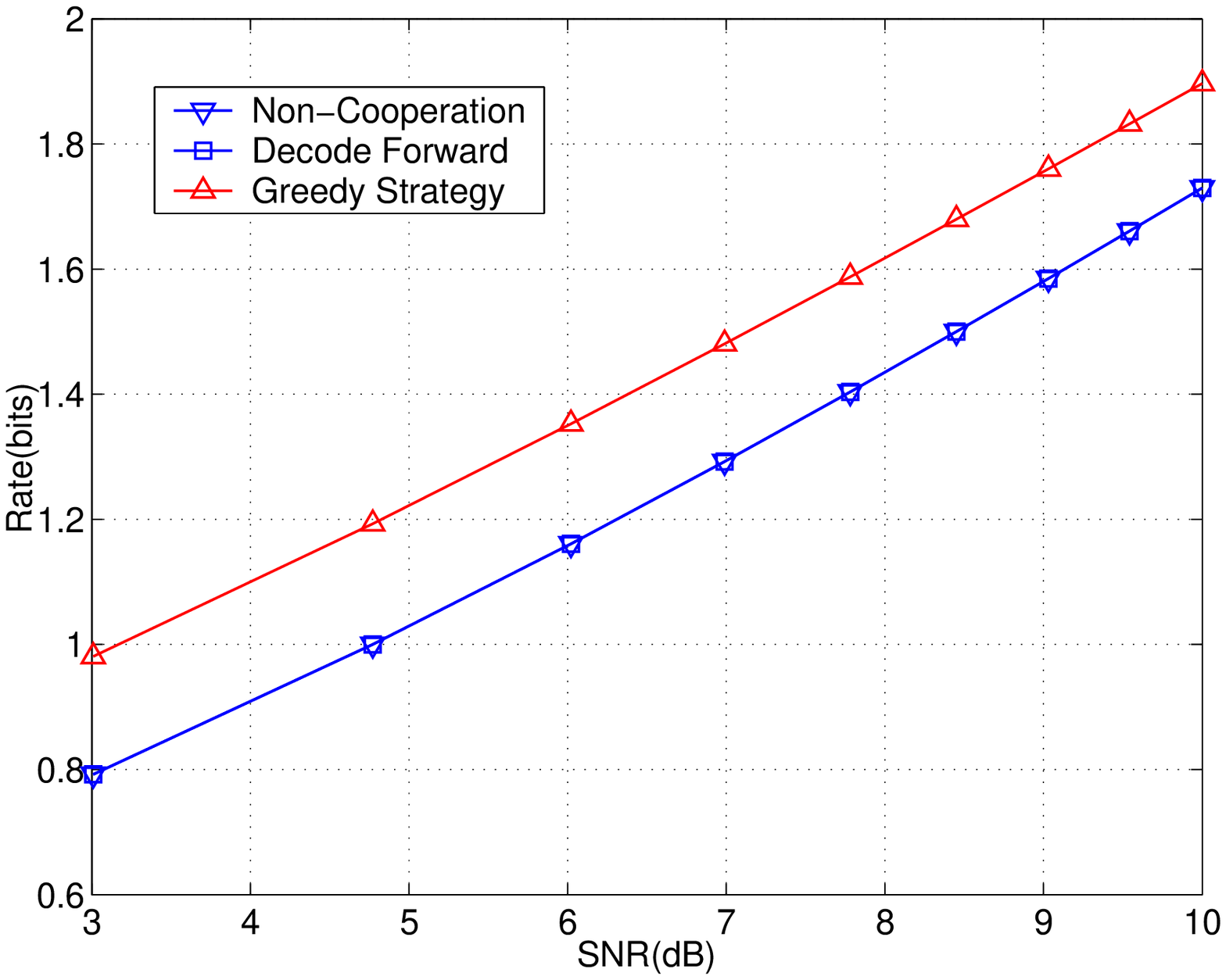}
\caption{The achievable rate of various schemes in the multicast
channel, $h_{12}=1$, $h_{13}=1$, and $h_{23}=23dB$. }
\label{fig:multicastpchangeh12small}
\end{center}
\end{figure}

\begin{figure}[thb]
\begin{center}
\includegraphics[width=0.7 \textwidth]{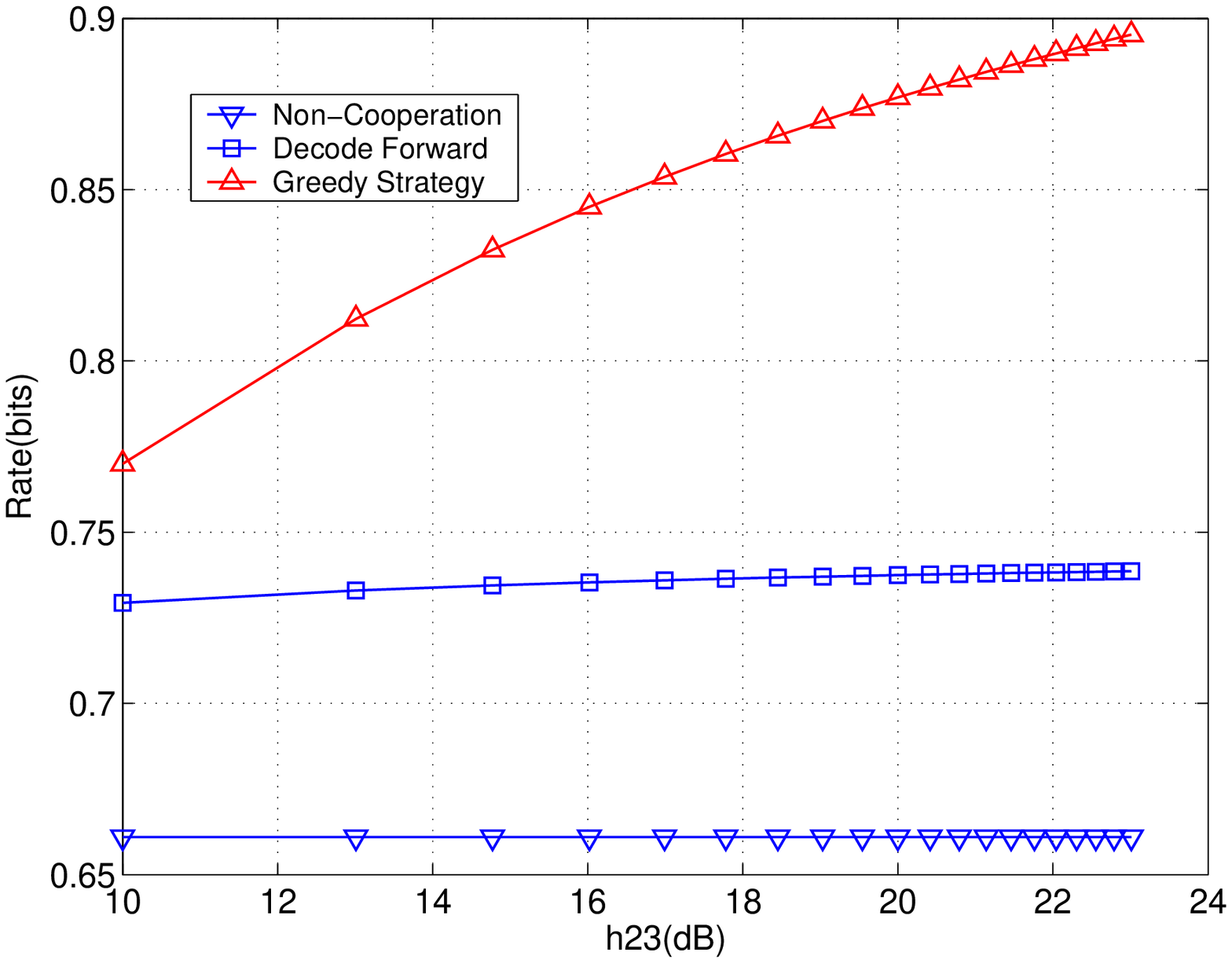}
\caption{The achievable rate of various schemes in the multicast
channel, $h_{12}=1.1$, $h_{13}=1$, and $\text{SNR}=1.8dB$. }
\label{fig:multicasth23change}
\end{center}
\end{figure}

Figure~\ref{fig:multicastpchange} compares the achievable rate of
the various multicast schemes where the DF cooperation strategy is
shown to outperform the non-cooperation scheme. It is also shown
that optimizing the parameter $\alpha$ provides an additional gain
(Note $R_{DF}$ in the figure corresponds to $\alpha=1$).
Figure~\ref{fig:multicastpchangeh12small} reports the achievable
rate of the three schemes when $h_{12}=h_{13}$. In this case, it
is easy to see that DF strategy yields {\bf exactly} the same
performance as the non-cooperative strategy. On the other hand, as
illustrated in the figure, the proposed greedy strategy is still
able to offer a sizable gain. Figure~\ref{fig:multicasth23change}
illustrates the fact that the gain of greedy strategy increases as
$h_{23}$ increases. The non-cooperation scheme is not able to
exploit the inter-receiver channel, and hence, its achievable rate
corresponds to a flat line. The DF scheme can benefit from the
inter-receiver channel, but its maximum rate is limited by
$C(h_{12}^2P)$, whereas the greedy strategy achieves a rate
$R_g=C((h_{12}^2+h_{13}^2)P)$ as $h_{23}\rightarrow \infty$.

\section{The Conference channel}\label{sec:conference}
Arguably the most demanding instantiation of the three-node
network is the conference channel. Here, the three nodes are
assumed to observe correlated data streams and every node is
interested in communicating its observations to the other two
nodes. In a first step to understand this channel, one is
naturally led to applying cut-set arguments to obtain a lower
bound on the necessary bandwidth expansion factor\footnote{Here,
we use the bandwidth expansion factor, instead of the achievable
rate, since it enjoys a nice additive property that will simplify
the development.}. To satisfy the conference channel requirements,
every node needs to transmit its message to the other two nodes
and receive their messages from them. Due to the half duplex
constraint, these two tasks cannot be completed simultaneously.
Take node-$1$ as an example and consider the transmission of a
block of observations $S_{1}^K$ to the other two nodes using
$N_{t}$ channel uses. To obtain a lower bound on the bandwidth
expansion factor, we assume that node-$2$ and node-$3$ can fully
cooperate, from a joint source-channel coding perspective, which
converts the problem into a point-to-point situation. Then
node-$1$ only needs to randomly divide its source sequences into
$2^{KH(S_{1}|S_{2},S_{3})}$ bins and transmit the corresponding
bin index. With $N_{t}$ channel uses, the information rate is
$\frac{KH(S_{1}|S_{2},S_{3})}{N_{t}}$. The channel capacity
between node-$1$ and the multi-antenna node-$2,3$ is
$C((h_{12}^2+h_{13}^2)P)$. In order to decode $S_{1}^K$ at
node-$2,3$ with a vanishingly small error probability, the
following condition must be satisfied,
\begin{equation}
\frac{KH(S_{1}|S_{2},S_{3})}{N_{t}}\leq
C((h_{12}^2+h_{13}^2)P).\nonumber
\end{equation}
Similarly, with full cooperation between node-$2$ and node-$3$,
the following condition is needed to ensure the decoding of the
sequence $S_{2}^K,S_{3}^K$ at node-$1$ with a vanishingly small
error probability,
\begin{equation}
\frac{KH(S_{2},S_{3}|S_{1})}{N_{r}}\leq
C((h_{12}^2+h_{13}^2)P).\nonumber
\end{equation}
These two genie-aided bounds at node-$1$ imply that the minimum
bandwidth expansion factor required for node-$1$ is
$\tau_{1,gen}=\frac{H(S_{1}|S_{2},S_{3})
+H(S_{2},S_{3}|S_{1})}{C((h_{12}^2+h_{13}^2)P)}$. Similarly, we
can obtain the corresponding genie-aided bounds for node-$2$ and
node-$3$, $\tau_{2,gen}=\frac{H(S_{2}|S_{1},S_{3})
+H(S_{1},S_{3}|S_{2})}{C((h_{12}^2+h_{23}^2)P)}$,
$\tau_{3,gen}=\frac{H(S_{3}|S_{1},S_{2})
+H(S_{1},S_{2}|S_{3})}{C((h_{13}^2+h_{23}^2)P)}$. To satisfy the
requirement for all these three nodes, the minimum bandwidth
expansion factor for this half-duplex conference channel is
therefore
\begin{equation}\label{genie}
\tau_{gen}\geq \max\limits_{i=1,2,3}\tau_{i,gen}.
\end{equation}

At this point, we remark that it is not clear whether the
genie-aided bound in (\ref{genie}) is achievable. Moreover,
finding the optimal cooperation strategy for the conference
channel remains an elusive task\footnote{Most of the earlier works
on the conference channel has focused on the source coding aspect,
e.g., \cite{Imre.source}.}. However, inspired by our greedy
multicast strategy, we propose in the following a modular
cooperation approach composed of three {\em cooperative multicast
with side information} stages. In this scheme, each node takes a
turn to multicast its information to the other two nodes. The
multicast problem here is more challenging than the scenario
considered in Section~\ref{sec:multicast} due to the presence of
correlated, and different, side information at the two receive
nodes. As argued in the following section, in order to fully
exploit this side information, one must adopt a cooperative
source-channel coding approach in every multicast stage.
Furthermore, from one stage to the next, the side-information
available at the different nodes changes. For instance, assuming
the first stage is assigned to node-$1$, then the side-information
available at node-$2$ and $3$ will enlarge after the first stage
to $(S_1,S_2)$ and $(S_1,S_3)$, respectively. Now, suppose
node-$2$ is scheduled to multicast next, then the rate required by
node-$3$ is now reduced to $H(S_2|S_1,S_3)$, thanks to the
additional side-information $S_1$. Thus, one can see that the
overall performance depends on the efficiency of the scheduling
algorithm. In Section~\ref{scheduler}, we present a greedy
scheduling algorithm that enjoys a low computational complexity
and still achieves a near-optimal performance.

\subsection{Multicast with Side-information}
The relay and multicast scenarios considered previously share the
common feature of the presence of only one source $S_1$. Here, we
expand our investigation of the multicast scenario by allowing the
receive nodes to observe correlated, and possibly different, side
information. To simplify the presentation, without sacrificing any
generality, we assume that node-$1$ is the source and node-$2$ and
$3$ are provided with the side information $S_2$ and $S_3$,
respectively. Before presenting our greedy cooperation strategy,
we study the non-cooperative scenario where the two receive nodes
are not allowed to communicate. This investigation yields an upper
bound on the bandwidth expansion factor achievable through
cooperation.

One can obtain a simple-minded transmission scheme by separating
the source and channel coding components. In this approach, by
appealing to the standard random binning argument
\cite{Imre.source}, node-$1$ encodes the source sequence at the
rate $\max\{H(S_1|S_2),H(S_1|S_3)\}$ which allows both receivers
to decode with the aid of their side-information. Such a source
code is then sent via the multicast channel assuming no receiver
cooperation, which corresponds to a transmission rate
$\min\{C(h_{12}^2P),C(h_{13}^2P)\}$. Therefore, the above scheme
would require
\begin{equation} \label{equ:naivemulticast}
\tau=\frac{\max\{H(S_1|S_2),H(S_1|S_3)\}}{\min\{C(h_{12}^2P),C(h_{13}^2P)\}}
\end{equation}
channel uses per source symbol. In \eqref{equ:naivemulticast}, the
source code based on random binning reflects the worst-case
scenario corresponding to the least correlated receiver node,
i.e., $\max\{H(S_1|S_2),H(S_1|S_3)\}$. A more efficient solution
utilizes a {\em nested binning} approach that combines the
information required by the two receive nodes into a single
hierarchical binning scheme. For the convenience of exposition, we
assume that $H(S_1|S_2)>H(S_1|S_3)$. A source sequence $s_1^K$ is
randomly assigned to one of $2^{KH(S_1|S_2)}$ bins. This is the
low-level indexing sufficient for node-$2$ to decode with
side-information $S_2$. These indices are then (randomly) divided
into $2^{KH(S_1|S_3)}$ equal-sized groups, which corresponds to
the random binning approach for node-$3$. Therefore, a source
sequence $s_1^K$ is associated with an index-pair $(b,c)$, where
$b\in[1,2^{KH(S_1|S_3)}]$ is the group index and
$c\in[1,2^{K(H(S_1|S_2)-H(S_1|S_3))}]$ identifies the bin index
within a group. Given side-information $S_3$ (more correlated with
the source), node-$3$ needs only the group index $b$ to recover
the source sequence. But the low-level bin index is necessary for
node-$2$ to decode. In summary, the above nested binning scheme
permits the source node to send $(b,c)$ to node-$2$ while only $b$
to node-$3$. Such a structured message is called the {\em degraded
information set} in \cite{Korner.broaddegrade} where $b$ is the
``common'' information for both receivers and $c$ the ``private''
information required by only one of the two receivers. The
corresponding rate set can be written as $(R_2,0,R_0)$, where
$R_2$ is the rate associated with the private message for
node-$2$, $R_0$ is the rate associated with the common message,
and node-$3$ receives no private information. This broadcast
channel with a degraded message set has been studied in
\cite{Korner.broaddegrade} (and references wherein).

\begin{thm}[see \cite{Korner.broaddegrade}] \label{them:broadwithdegra}
The capacity region of broadcast with degraded information set is
the convex hull of the closure of all $(R_{2},0,R_{0})$ satisfying
\begin{equation}
\begin{split}
R_{2}&\leq I(X_{1};Y_{2}|U),\\
R_{0}&\leq I(U;Y_{3}),\\
R_{2}+R_{0}&\leq I(X_{1};Y_{2}),
\end{split}
\end{equation}
for some joint distribution $p(u)p(x_{1}|u)p(y_{2},y_{3}|x_{1})$.
\end{thm}

If the broadcast channel itself is degraded then the above three
constraints can be simplified \cite{Korner.broaddegrade}. Consider
the case where $X_1\rightarrow Y_3\rightarrow Y_2$ forms a Markov
chain. If the last constraint is satisfied (in which case node-$2$
can decode both the private and common message), node-$3$ can also
decode both parts of the source message although it does not need
the private part. On the other hand, if $X_1\rightarrow
Y_2\rightarrow Y_3$, the problem can be reduced to the
conventional (degraded) broadcast setting where the rate
$R_2$/$R_0$ is for node-$2$/$3$ and node-$2$ would automatically
decode the message for the degraded node-$3$. So, in this case,
only the first two constraints are sufficient. The final step, in
this approach, is to combine Theorem~\ref{them:broadwithdegra}
with our nested binning approach. In this case, the rate set is
given by $R_2=\frac{K(H(S_1|S_2)-H(S_1|S_3))}{N}$ and
$R_0=\frac{KH(S_1|S_3)}{N}$ and we obtain the following result.

\begin{lem}\label{lem:multicastsidedegraded}
For multicast with side-information, the achievable bandwidth
expansion factor $\tau=N/K$ based on nested binning source coding
and degraded information set broadcasting is given by
\begin{enumerate}
\item if $h_{12}^2<h_{13}^2$,
\begin{equation}
H(S_{1}|S_{2})\leq \tau C(h_{12}^2P).
\end{equation}
\item if $h_{12}^2>h_{13}^2$,
\begin{equation}
\begin{split}
&H(S_{1}|S_{2})-H(S_{1}|S_{3})\leq \tau C\bigg(\gamma h_{12}^2P\bigg),\\
&H(S_{1}|S_{3})\leq \tau
C\bigg(\frac{(1-\gamma)h_{13}^2P}{1+\gamma h_{13}^2P}\bigg).
\end{split}
\end{equation}
for some $\gamma$.
\end{enumerate}
\end{lem}

Now, we are ready to describe our greedy cooperative
source-channel coding approach. Similar to the multicast scenario,
the receive nodes follow a greedy strategy to determine the order
of decoding. Due to the presence of side information, however, a
more careful approach must be employed in choosing the {\em
strong} receiver. To illustrate the idea, consider the following
degenerate case, where $S_{3}\;\text{is independent of}\;S_{1}$,
$S_{1}=S_{2}$, and $h_{12}^2<h_{13}^2$.  Although the channel
between node-$1$ and node-$2$ is worse in this case, node-$2$
knows the information $S_{1}^K$ from the beginning because
$S_{1}=S_{2}$. So it can start to cooperate with node-$1$ from the
very beginning. This toy example suggests that one should take the
amount of side information available at each node into
consideration. In our scheme, each node calculates the expected
bandwidth expansion factor assuming no receiver cooperation,
$\tau_{ex,i}=H(S_1|S_i)/C_i$, where $C_i$ denotes the link
capacity $C(h_{1i}^2P)$. The receive node with the smaller
$\tau_{ex}$ is deemed as the {\em strong} node, and hence, will
decode first. Our definition of strong and weak highlights the
cooperative source-channel coding approach proposed in this paper.
Without loss of generality, we assume $\tau_{ex,2}<\tau_{ex,3}$.
However, the ``weak'' node-$3$ may still decide to assist node-$2$
in decoding through a CF approach, in a way similar to
Section~\ref{sec:multicast}, hoping to benefit from node-$2$'s
help after it decodes. After node-$2$ successfully decoding,
with/without the additional help from node-$3$, it coordinates
with the source node to facilitate decoding at node-$3$, in order
to start the next round of multicast.

To better describe the cooperative source-channel coding, we
consider first the simple case where node-$3$ does not help
node-$2$. We randomly bin the sequences $S_{1}^K$ into
$2^{KH(S_{1}|S_{2})}$ bins and denote the bin index by
$w\in[1,2^{KH(S_1|S_2)}]$. We further denote by $f_{s1}$ the
mapping function $w=f_{s1}(s_{1}^K)$. We then independently
generate another bin index $b$ for every sequence $S_{1}^K$ by
picking $b$ uniformly from $\{1,2,\ldots,2^{KR}\}$, where $R$ is
to be determined later. Let $B(b)$ be the set of all sequences
$S_{1}^K$ allocated to bin $b$. Thus, every source sequence has
two bin indices $\{w,b\}$ associated with it. A full cooperation
cycle is divided into two stages, where we refer to the network
state in these two stages as $m_{1}$ and $m_{2}$, respectively. In
the first stage using for $N_{1}$ channel uses, node-$1$ sends the
message $w$ to node-$2$ using a capacity achieving code. This
stage is assumed to last for $N_{1}$ channel uses. At the end of
this state, node-$2$ can get a reliable estimate $\hat{w}=w$ if
the condition $KH(S_1|S_2)\leq N_1C(h_{12}^2P)$ is satisfied.
Next, node-$2$ searches in the bin specified by $\hat{w}$ for the
one and only one $\hat{s}_{21}^K$ that is typical with $s_2^K$. If
none exists, decoding error is declared, otherwise,
$\hat{s}_{21}^K$ is the decoding sequence. During this stage,
node-$3$ computes a list $\ell(\mathbf{y}_{3,m_1})$ such that if
$w'\in \ell(\mathbf{y}_{3,m_{1}})$ then
$\{\mathbf{x}_{1,m_{1}}(w'),\mathbf{y}_{3,m_{1}}\}$ are jointly
typical. A key point of our scheme is that node-$3$ does not
attempt to decode $w$, but rather proceeds to decoding $s_{1}^K$
directly. After node-$2$ decodes $s_{1}^K$ correctly, it knows the
pair $\{w,b\}$, and hence, in the second stage node-$2$ and
node-$1$ cooperate to send the message $b$ to node-$3$. At the end
this stage, if the parameters are appropriately chosen, node-$3$
can decode $b$ correctly. Node-$3$ then searches in the bin $B(b)$
for the one and only one $\hat{s}_{31}^K$ that is jointly typical
with $s_{3}^K$ and that $f_{s1}(\hat{s}_{31}^K)\in
\ell(\mathbf{y}_{3,m_{1}})$.

\begin{lem}\label{lemma:multicastsideinfor}
With the proposed scheme, both node-$2,3$ can decode $S_{1}^K$
with a vanishingly small probability of error if
$\tau_{0}=N_{0}/K,\tau_{1}=N_{1}/K$ satisfy the following
conditions
\begin{equation}
\begin{split}
H(S_{1}|S_{2})&\leq \tau_{1} C(h_{12}^2P),\\
H(S_{1}|S_{3})-\frac{\min\{C(h_{13}^2P),C(h_{12}^2P)\}
H(S_{1}|S_{2})}{C(h_{12}^2P)}&\leq
\tau_{0}C((h_{13}^2+h_{23}^2)P).
\end{split}
\end{equation}
\end{lem}
\begin{proof}
Please refer to Appendix~\ref{appex:multiside}.
\end{proof}
Next, we allow for the weak node-$3$ to assist the strong node-$2$
in decoding. The original $N_1$ channel uses now split into two
parts: 1) state $m_1$ occupying $\alpha N_1$ channel uses during
which both receiver nodes listen to the source node; and 2) state
$m_3$ of the remaining $(1-\alpha)N_1$ channel uses during which
node-$3$ sends a compressed received signal to node-$2$. At the
end of the $N_1$ network uses, node-$2$ decodes the source
sequence and then proceeds to facilitate the same list-decoding at
the other receiver as described above. The simple case where
node-$3$ does not assist node-$2$ can be regarded as a special
case of the greedy scheme when $\alpha=1$. Slightly modifying the
proof of Lemma~\ref{lemma:multicastsideinfor}, we obtain

\begin{lem}\label{lem:multicastsidefb}
If $\tau_{0},\tau_{1}$ satisfy the following conditions, both
node-$2,3$ will decode $S_{1}^K$ with vanishingly small
probability of error.
\begin{equation}
\begin{split}
H(S_{1}|S_{2})&\leq \tau_{1} R_{CF2}(\alpha),\\
H(S_{1}|S_{3})-\frac{\alpha
\min\{I(X_{1};Y_{3}|m_{1});I(X_{1};\hat{Y}_{3},Y_{2}|m_{1})\}
H(S_{1}|S_{2})}{R_{CF2}(\alpha)}&\leq
\tau_{0}C((h_{13}^2+h_{23}^2)P).
\end{split}
\end{equation}
Where $R_{CF2}(\alpha)$ is the achievable rate of compress-forward
scheme for the following relay channel: node-$1$ acts as the
source, node-$3$ the relay that spends $1-\alpha$ part of the time
 in helping destination using CF scheme, and node-$2$ the destination.
 The symbol $\hat{Y}_3$ stands for the compressed version of the received
 signal at node-$3$ ($Y_3$).
\end{lem}

Unfortunately, the expressions for the achievable bandwidth
expansion factors do not seem to allow for further analytical
manipulation. In order to shed more light on the relative
performance of the different schemes, we introduce the {\em
minimum energy per source observation} metric. Given the total
transmission power $P$, the bandwidth expansion factor $\tau$
translates to the energy requirement per source observation as
\begin{equation}
E(P)=\tau(P) P=\frac{N(P)P}{K}.
\end{equation}
Let $E_1(P)$ denotes the energy per source symbol for the
benchmark based on broadcast with a degraded information set and
$E_2(P)$  for the proposed cooperative multicast scheme. It is
easy to see that both $E_1(P)$ and $E_2(P)$ are non-increasing
function of $P$, and hence, approach their minimal values as
$P\rightarrow 0$, that is
\begin{equation}
E_{i,m}=\lim_{P\rightarrow 0} E_{i}(P) \mbox {  for  } i
\in\{1,2\}.
\end{equation}

Under the assumption that $\tau_{ex,2}<\tau_{ex,3}$ and using
Lemmas~\ref{lem:multicastsidedegraded}~and~\ref{lem:multicastsidefb},
one obtain:

\begin{thm}\label{them:multisideener}
\begin{enumerate}
\item Broadcast with degraded information set:\newline When
$H(S_{1}|S_{2})>H(S_{1}|S_{3})$,
\begin{equation}
E_{1,m}=\frac{2}{\log
e}\bigg(\frac{H(S_{1}|S_{2})}{h_{12}^2}+\big(\frac{1}{h_{13}^2}
-\frac{1}{h_{12}^2}\big)^{+}H(S_{1}|S_{3})\bigg).
\end{equation}
When $H(S_{1}|S_{3})>H(S_{1}|S_{2})$,
\begin{equation}
E_{1,m}=\frac{2}{\log
e}\bigg(\frac{H(S_{1}|S_{3})}{h_{13}^2}+\big(\frac{1}{h_{12}^2}
-\frac{1}{h_{13}^2}\big)^{+}H(S_{1}|S_{2})\bigg).
\end{equation}
Here $x^+=\max\{x,0\}$. \item Greedy strategy:
\begin{equation}
E_{2,m}=\frac{2}{\log
e}\bigg(\frac{H(S_{1}|S_{2})}{h_{12}^2}+\frac{h_{12}^2H(S_{1}|S_{3})
-\min\{h_{13}^2,h_{12}^2\}H(S_{1}|S_{2})}{(h_{13}^2+h_{23}^2)h_{12}^2}\bigg).
\end{equation}
\item $E_{2,m}<E_{1,m}$.
\end{enumerate}
\end{thm}
\begin{proof}
Please refer to Appendix~\ref{appex:multisideener}.
\end{proof}

\begin{figure}[htb]
\begin{center}
\includegraphics[width=0.7\textwidth]{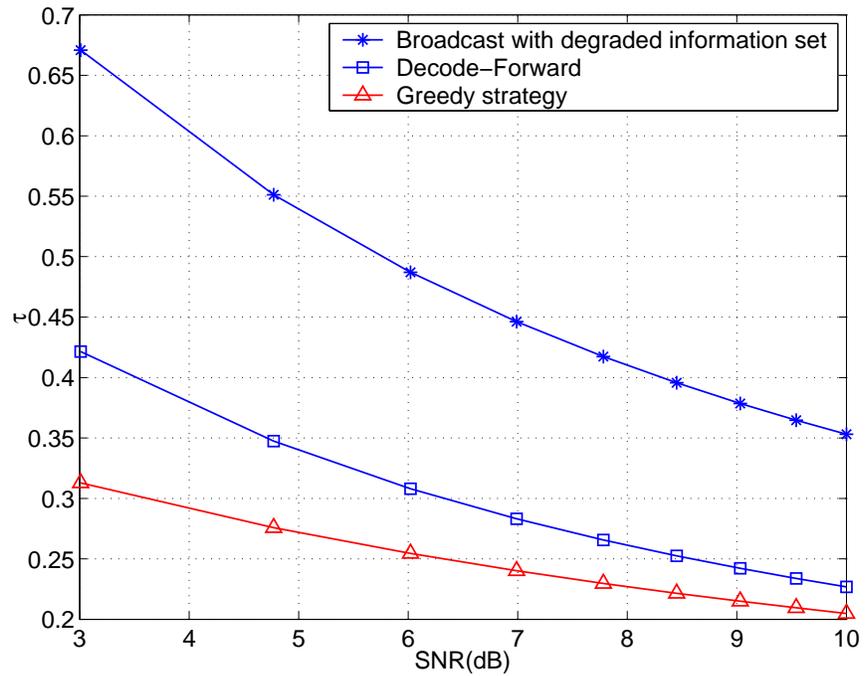}
\caption{The bandwidth expansion factor of various schemes in the
multicast channel with side-information, $h_{12}=2,h_{13}=1,
h_{23}=90,H(S_{1}|S_{2})=0.9, H(S_{1}|S_{3})=0.3$.}
\label{fig:mulsidebandexpansion}
\end{center}
\end{figure}

\begin{figure}[htb]
\begin{center}
\includegraphics[width=0.7 \textwidth]{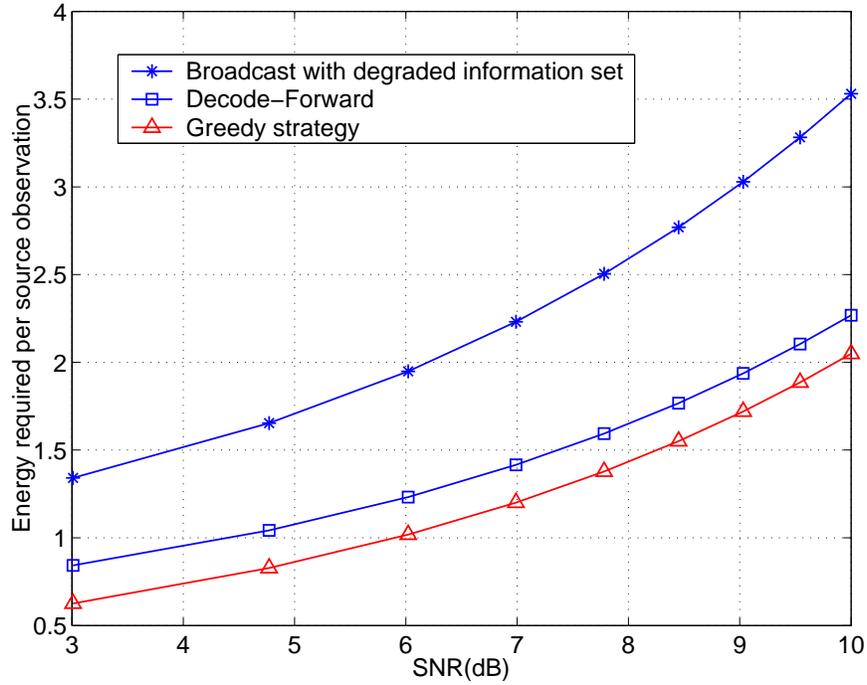}
\caption{ The energy required per source observation of various
schemes in the multicast channel with side-information,
$h_{12}=2,h_{13}=1,h_{23}=90,H(S_{1}|S_{2})=0.9,
H(S_{1}|S_{3})=0.3$.} \label{fig:mulsideenergy}
\end{center}
\end{figure}

In general, Theorem~\ref{them:multisideener} reveals the
dependence of the minimum energy per source observation for the
two schemes on the correlation among the source variables and the
channel gains. However, Theorem~\ref{them:multisideener} also
establishes the superiority of the cooperative source-channel
coding over the non-cooperative benchmark in the general case (at
least with respect to the minimum energy requirement). Thus, when
combined with the results in Section~\ref{sec:multicast}, this
result argues strongly for receiver cooperation in the multicast
scenario even under the stringent half-duplex and total power
constraints. Finally,
Figures~\ref{fig:mulsidebandexpansion}~and~\ref{fig:mulsideenergy}
validate our theoretical claims.
Fig.~\ref{fig:mulsidebandexpansion} reports the relationship
between power and bandwidth expansion factor, whereas
Fig.~\ref{fig:mulsideenergy} reports the relationship between
power and energy per source observation for various schemes. In
both figures, the gain offered by the proposed cooperative
multicast scheme is evident.

\subsection{Multicast scheduler}\label{scheduler}
The second step in the proposed solution for the conference
channel is the design of the scheduler. As argued earlier, the
efficiency of the scheduler has a critical impact on the overall
performance. Given a specific multicast order, one can compute the
overall bandwidth expansion factor by adding up the required
$\tau$ for every multicast stage. The {\em optimal} scheduler will
choose the multicast order corresponding to the minimum bandwidth
expansion factor among all possible permutations. The following
result argues for the efficiency of our proposed cooperation
scheme for the conference channel.
\begin{thm}\label{thm:conference}
The cooperative source-channel coding multicast scheme with the
optimal scheduler has the following properties (in the conference
channel):
\begin{enumerate}
\item It is asymptotically optimal, i.e., achieves the genie-aided
bound, when any one of the channel coefficients is sufficiently
large. \item It always outperforms the broadcast with a degraded
set based multicast scheme with the optimal scheduler.
\end{enumerate}
\end{thm}
\begin{proof}
Please refer to Appendix~\ref{appex:conference}.
\end{proof}

One can argue, however, that the optimal scheduler suffers from a
high computation complexity since every node is required to
compute the overall bandwidth expansion factors for the six
possible scheduling alternatives. To reduce the computational
complexity, one can adopt the following greedy strategy. At the
beginning of every multicast stage, every node that has not
finished multicasting yet will calculate its expected bandwidth
expansion factor based on the cooperative scheme for multicast
with side-information. The greedy scheduler chooses the node with
the least expected bandwidth expansion factor to transmit at this
stage. After this node finishes and the side-information is
updated, the scheduler computes the expected bandwidth expansion
factor for the rest of nodes and selects the one with the least
bandwidth expansion factor to multicast next. As a side comment,
we note here that this approach is easily scalable for networks
with more than three nodes. In general, this greedy scheduler
constitutes a potential source for further sub-optimality.
However, it can achieve the genie-aided bound in the asymptotic
limit when one of the channel gains is sufficiently large. Take
$h_{23}\rightarrow \infty$ as an example. In this case, if one of
the following conditions is satisfied, the greedy scheduler
achieves the genie-aided bound:
\begin{enumerate}

\item
$H(S_{2}|S_{1})<\min\{H(S_{1}|S_{2}),H(S_{1}|S_{3}),H(S_{3}|S_{1})\}\;\text{and
}\;H(S_{3}|S_{1},S_{2})<H(S_{1}|S_{2},S_{3})$,

\item
$H(S_{3}|S_{1})<\min\{H(S_{1}|S_{2}),H(S_{1}|S_{3}),H(S_{2}|S_{1})\}\;\text{and
}\;H(S_{2}|S_{1},S_{3})<H(S_{1}|S_{2},S_{3})$.
\end{enumerate}
One can easily check that if 1) is satisfied, the greedy scheduler
will choose the sequence $2\rightarrow 3\rightarrow 1$ as the
multicast order, which is the optimal order that achieves the
genie-aided bound. If 2) is satisfied, the greedy scheduler will
choose the sequence $3 \rightarrow 2\rightarrow 1$ as the
multicast order, which is also the order that achieves the
genie-aided bound in this case.

\begin{figure}[htb]
\begin{center}
\includegraphics[width=0.7 \textwidth]{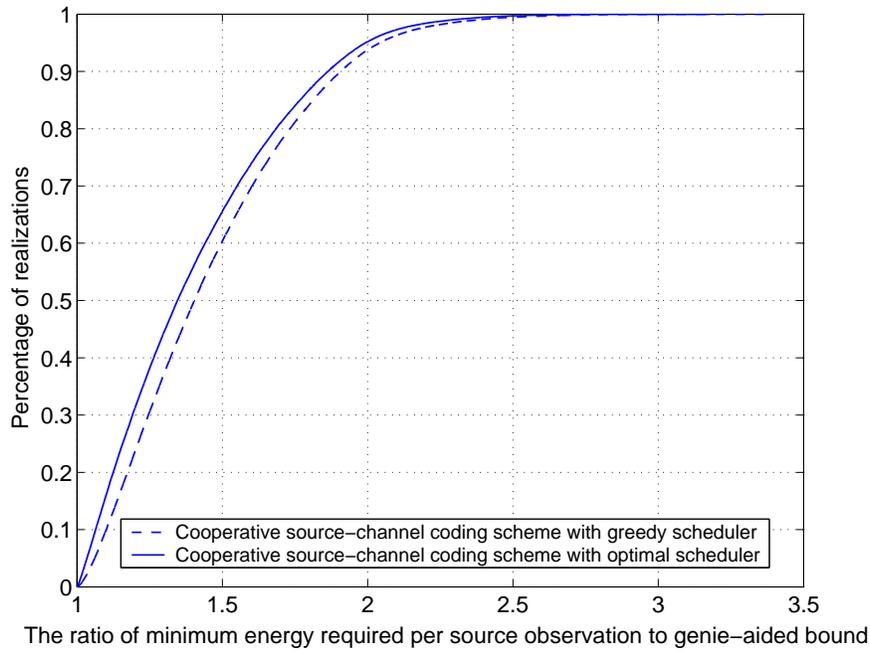}
\caption{The ratio of the energy required per source observation
of the cooperative source-channel coding scheme to the genie aid
bound.} \label{fig:coopcdf}
\end{center}
\end{figure}

\begin{figure}[htb]
\begin{center}
\includegraphics[width=0.7\textwidth]{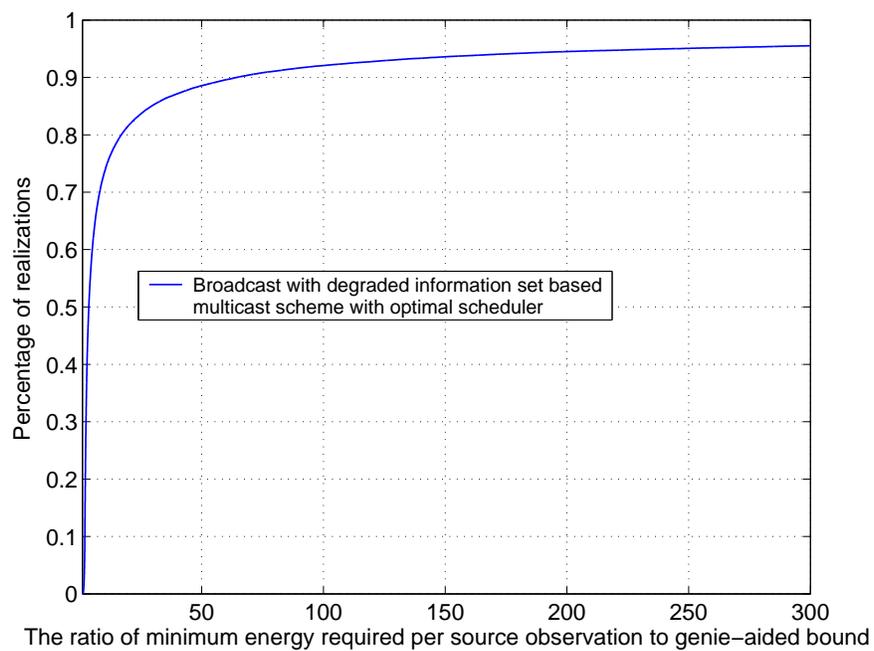}
\caption{The ratio of the energy required per source observation
of the broadcast with degraded information set based multicast
with optimal scheduler to the genie aid bound.}
\label{fig:noncopcdf}
\end{center}
\end{figure}

The numerical results in
Figures~\ref{fig:coopcdf}~and~\ref{fig:noncopcdf} validate our
claims on the efficiency of the proposed cooperation strategy.
These figures compare the minimum energy required per source
observation by each scheme, under randomly generated channel gains
and correlation patterns. For each realization, we use numerical
methods to find the optimal order and greedy order for cooperative
scheme, and the corresponding minimum energy required per source
observation, namely $E_{oc},E_{gc}$. We also find the optimal
order for the non-cooperative scheme and the corresponding minimum
energy required per source observation $E_{nc}$. The minimum
energy required per source observation by the genie-aided bound
$E_{gen}$ is used as a benchmark. In particular, for each
realization, we calculate the ratio of the minimum energy required
by the three schemes to the genie aided bound. We repeat the
experiment $100000$ times and report the histogram of the ratios
in the figures. In Fig.~\ref{fig:coopcdf}, we see that $94$
percent of the time, the proposed cooperative scheme with the
greedy scheduler operates within $3$ dB of the genie-aided bound.
We also see that the performance of greedy scheduler is almost
identical to the optimal scheduler. Figure~\ref{fig:noncopcdf}
shows that the non-cooperative scheme operates more than $3$ dB
away from the genie-aided bound for $90$ percent of the time.
Moreover, there is a non-negligible probability, i.e., $8\;
\text{percent}$, that this scheme operates $100$ dB away from
genie aided bound. It is clear that receiver cooperation reduces
this probability significantly.

\section{Conclusions}\label{sec:conclusion}
We have adopted a formulation of the three-node wireless network
based on the half-duplex and total power constraints. We argued
that this formulation unifies many of the special cases considered
in the literature and highlights their structural similarities. In
particular, we have proposed a greedy cooperation strategy in
which the {\em weak} receiver first helps the {\em strong}
receiver to decode in a CF configuration. After successfully
decoding, the strong user starts assisting the weak user in a DF
configuration. We have shown that different instantiations of this
strategy yield excellent performance in the relay channel with
feedback, multicast channel, and conference channel. Our analysis
for the achievable rates in such special cases sheds light on the
value of feedback in relay channel and the need for a cooperative
source-channel coding approach to efficiently exploit receiver
side information in the wireless setting.

Extending our work to networks with arbitrary number of nodes
appears to be a natural next step. In particular, the
generalization of the greedy cooperation strategy is an
interesting avenue worthy of further research. Our preliminary
investigations reveal that such a strategy can get sizable
performance gains, over the traditional multi-hop routing
approach, in certain network configurations.



%
%

\appendices

\section{Proof of Lemma \ref{lem:fb}}\label{appex:fb}
To facilitate exposition we first prove the result for the
discrete memoryless channel (DMC) and then progress to the
Gaussian channel. In this paper, we refer to typical sequences as
strong typical sequcences (see~\cite{T.Cover, T.Cover1,
Kramer.relay} for details of strong typical sequences).

\subsection{Discrete Memoryless Channel}
\subsubsection{Outline}
Suppose we want to send i.i.d.\ source $w(i),w(i)\in[1,M]$, in
which $M=2^{N R}$ to destination. Equally divide these $2^{NR}$
messages into $M_{1}=2^{N\alpha tR_{1}}$ cells, index the cell
number as $b(i)$. Index the element in every cell as $d(i),
d(i)\in [1,M_{2}], M_{2}=2^{ N(1-t)\cdot R_{2}}$. Thus
\begin{equation}
2^{N R}=2^{N\alpha t R_{1}} 2^{ N(1-t)R_{2}}
\end{equation}
that is
\begin{equation}
R=\alpha tR_{1}+(1-t) R_{2}
\end{equation}
The main idea is that the relay and the destination help each
other to decode $b(i)$:
\begin{itemize}
\item In the first state $m_{1}$, source sends the cell index
$b(i)$ to both the relay and the destination. At this time,
neither the relay nor the destination can decode this information.
\item In the feedback state $m_{3}$, the destination sends
compressed version of the received noisy signal to the
destination. At the same time, the source sends additional
information to the relay. \item At the end of the relay receive
mode, the relay gets an estimation of $b(i)$, namely $\hat{b}(i)$.
Thus in $m_{2}$, the relay sends its knowledge of $\hat{b}(i)$ to
the destination to help it decode $b(i)$. At the same time, the
source sends $d(i)$ to the destination.
\end{itemize}
\subsubsection{Random code generation}
Fix
$p(x_{1}|m_{1}),p(x_{1},x_{3}|m_{3}),p(x_{1},x_{2}|m_{2}),p(\hat{y}_{3})$.

\begin{itemize}
\item state $m_{1}$:\newline At the source, generate $2^{\alpha
tNR_{1}}$ i.i.d.\ length-$\alpha tN$ sequence
$\mathbf{x}_{1,m_{1}}$ each with probability
$p(\mathbf{x}_{1,m_{1}})=\prod \limits_{j=1}^{\alpha
tN}p(x_{1j}|m_{1})$. Label these sequences as
$\mathbf{x}_{1,m_{1}}(b)$, where $b\in[1,2^{\alpha tNR_{1}}]$ is
called the cell index.

\item state $m_{3}$:
\begin{itemize}
\item source node:\newline Generate $2^{(1-\alpha)tNR_{4}}$
i.i.d.\ length-$(1-\alpha)tN$ codewords $\mathbf{x}_{1,m_{3}}$
with $p(\mathbf{x}_{1,m_{3}})=\prod
\limits_{j=1}^{(1-\alpha)tN}p(x_{1j}|m_{3})$. Label these
sequences as $\mathbf{x}_{1,m_{1}}(q),
q\in[1,2^{(1-\alpha)tNR_{4}}]$. Randomly partition the $2^{\alpha
tNR_{1}}$ cell indices $\{b\}$ into $2^{(1-\alpha)tNR_{4}}$ bins
$Q_{q}$ with $q\in[1,2^{(1-\alpha)tNR_{4}}]$.

\item destination node:\newline Generate $2^{(1-\alpha)tNR_{3}}$
i.i.d.\ length-$(1-\alpha)tN$ codewords $\mathbf{x}_{3,m_{3}}$
with $p(\mathbf{x}_{3,m_{3}})=\prod
\limits_{j=1}^{(1-\alpha)tN}p(x_{3j}|m_{3})$. Index them as
$\mathbf{x}_{3,m_{3}}(u)$. Generate $2^{\alpha tN\hat{R}}$ i.i.d.\
length-$\alpha tN$ sequences $\hat{\mathbf{y}}_{3}(z)$ with
$p(\hat{\mathbf{y}}_{3})=\prod\limits_{j=1}^{\alpha
tN}p(\hat{y}_{3j})$. Randomly partition the set $z\in[1,2^{\alpha
tN\hat{R}}]$ into bins $U_{u}$, $u\in[1,2^{(1-\alpha)tNR_{3}}]$.
\end{itemize}

\item state $m_{2}$:
\begin{itemize}
\item relay node: \newline Randomly generate
$M_{0}=2^{(1-t)NR_{0}}$ i.i.d.\ length-$(1-t)N$ sequences
$\mathbf{x}_{2,m_{2}}$ with $p(\mathbf{x}_{2,m_{2}})=\prod
\limits_{j=1}^{(1-t)N} p(x_{2j}|m_{2})$. Index them as
$\mathbf{x}_{2,m_{2}}(c),c\in [1, 2^{(1-t)NR_{0}}]$. Randomly
partition the $2^{\alpha tNR_1}$ cell indices into $2^{(1-t)NR_0}$
bins $C_c$.

\item source node:\newline Generate $M_{2}=2^{(1-t)NR_{2}}$
i.i.d.\ length-$(1-t)N$ sequences $\mathbf{x}_{1,m_{2}}$ with
$p(\mathbf{x}_{1,m_{2}})=\prod \limits_{j=1}^{(1-t)N}
p(x_{1j}|x_{2j},m_{2})$ for every $\mathbf{x}_{2,m_{2}}$ sequence.
Index them $\mathbf{x}_{1,m_{2}}(d|c),d\in [1,2^{(1-t)NR_{2}}]$.
\end{itemize}
\end{itemize}

\subsubsection{Encoding}
Partition the source message set into $2^{\alpha tNR_1}$
equal-sized cells. Let $w(i)$ be the message to be sent in block
$i$. Suppose $w(i)$ is the $d(i)$-th message in cell-$b(i)$ and
the cell index $b(i)$ is in bin-$q(i)$ and bin-$c(i)$
respectively. For brevity we drop the block index $i$ in the
following.

\begin{itemize}
\item state $m_{1}$: \newline The source sends
$\mathbf{x}_{1,m_{1}}(b)$.

\item state $m_{3}$:
\begin{itemize}
\item The source node knows that the cell index $b$ is in bin-$q$,
so it sends $\mathbf{x}_{1,m_{3}}(q)$.

\item The destination first selects $\hat{\mathbf{y}}_3(z)$ that
is jointly typical with $\mathbf{y}_{3,m_{1}}$. It then sends
$\mathbf{x}_{3,m_3}(u)$ where $z$ is in the bin $U_u$.
\end{itemize}

\item state $m_{2}$:
\begin{itemize}
\item Knowing the cell index $b$ is in bin-$c$, the source node
sends the corresponding $\mathbf{x}_{1,m_{3}}(d|c)$.

\item Using the information received in state $m_{1}$ and $m_{3}$,
the relay gets an estimation of the cell index $\hat{\hat{b}}$.
Suppose $\hat{\hat{b}}$ is in bin-$\hat{\hat{c}}$. Then it sends
$\mathbf{x}_{2,m_{3}}(\hat{\hat{c}})$.
\end{itemize}
\end{itemize}

\subsubsection{Decoding}
In the following, code length $N$ is chosen sufficiently large.

\begin{itemize}
\item at the end of $m_{1}$: \newline The destination has received
$\mathbf{y}_{3,m_{1}}$ and it decides a sequence
$\hat{\mathbf{y}}_{3}(z)$ if
$(\hat{\mathbf{y}}_{3}(z),\mathbf{y}_{3,m_{1}})$ are jointly
typical. There exists such a $z$ with high probability if
\begin{equation}\label{equ:rhat1}
\hat{R}\geq I(\hat{Y}_{3};Y_{3}).
\end{equation}

\item at the end of $m_{3}$: \newline At this stage, only the
relay decodes the message.

\begin{itemize}
\item The relay estimates $u$ by looking for the unique $\hat{u}$
such that $(\mathbf{x}_{3,m_{3}}(\hat{u}),\mathbf{y}_{2,m_{3}})$
are jointly typical. $\hat{u}=u$ with high probability if
\begin{equation}
R_{3}\leq I(X_{3};Y_{2}).
\end{equation}

\item Knowing $\hat{u}$, the relay tries to decode $q$ by
selecting the unique $\hat{q}$ such that\newline
$(\mathbf{x}_{1,m_{3}}(\hat{q}),
\mathbf{x}_{3,m_{3}}(\hat{u}),\mathbf{y}_{2,m_{3}})$ are jointly
typical. $\hat{q}=q$ with high probability if
\begin{equation}
R_{4}\leq I(X_{1};Y_{2}|X_{3}).
\end{equation}

\item The relay calculates a list $\ell(\mathbf{y}_{2,m_{1}})$
such that $z\in \ell(\mathbf{y}_{2,m_{1}})$ if
$(\hat{\mathbf{y}}_{3,m_{1}}(z),\mathbf{y}_{2,m_{1}})$ are jointly
typical. Assuming $u$ decoded successfully at the relay, $\hat{z}$
is selected if it is the unique $\hat{z}\in U_{u}\bigcap
\ell(\mathbf{y}_{2,m_{1}})$. Using the same argument as in
\cite{T.Cover}, it can be shown that $\hat{z}=z$ occurs with high
probability if
\begin{equation}\label{equ:rhat2}
\alpha t\hat{R}\leq \alpha
tI(\hat{Y}_{3};Y_{2}|m_{1})+(1-\alpha)tR_{3}.
\end{equation}

\item The relay computes another list
$\ell(\mathbf{y}_{2,m_{1}},\hat{\mathbf{y}}_{3,m_{1}})$ such that
$b\in \ell(\mathbf{y}_{2,m_{1}},\hat{\mathbf{y}}_{3,m_{1}})$
if\newline
$(\mathbf{x}_{1,m_{1}}(b),\mathbf{y}_{2,m_{1}},\hat{\mathbf{y}}_{3,m_{1}})$
are jointly typical.

\item Finally, the relay declares $\hat{\hat{b}}$ is received if
it is the unique $\hat{b}\in Q_{q}\bigcap
\ell(\mathbf{y}_{2,m_{1}},\hat{\mathbf{y}}_{3,m_{1}})$. Using the
same arguement as in \cite{T.Cover}, one can show
$\hat{\hat{b}}=b$ with high probability if
\begin{equation}\label{equ:rc1}
\alpha tR_{1}\leq \alpha
tI(X_{1};\hat{Y}_{3},Y_{2}|m_{1})+(1-\alpha)tI(X_{1};Y_{2}|X_{3},m_{3}).
\end{equation}
\end{itemize}

\item at the end of $m_{2}$:
\begin{itemize}
\item The destination declares that $\hat{c}$ was sent from the
relay if there exists one and only one $\hat{c}$ such that
$(\mathbf{x}_{2,m_{2}}(\hat{c}),\mathbf{y}_{3,m_{2}})$ are jointly
typical. Then $\hat{c}=c$ with high probability if
\begin{equation}\label{equ:four}
R_{0}\leq I(X_{2};Y_{3}|m_{2}).
\end{equation}

\item After decoding $\hat{c}$, the destination further declares
that $\hat{d}$ was sent from the source if it is the unique
$\hat{d}$ such that
$(\mathbf{x}_{1,m_{2}}(\hat{d}),\mathbf{x}_{2,m_{2}}(\hat{c}),\mathbf{y}_{3,m_{2}},)$
are joint typical. Assuming $c$ decoded correctly, the probability
of error of $\hat{d}$ is small if
\begin{equation}\label{equ:rq2}
R_{2}\leq I(X_{1};Y_{3}|X_{2},m_{2}).
\end{equation}

\item At first, the destination calculates a list
$\ell(\mathbf{y}_{3,m1})$, such that $b\in
\ell(\mathbf{y}_{3,m1})$ if
$(\mathbf{x}_{1,m_{1}}(b),\mathbf{y}_{3,m1})$ are jointly typical.
 Assuming $c$ decoded successfully at the destination, $\hat{b}$
is declared to be the cell index if there is a unique $\hat{b}\in
C_{c}\bigcap \ell(\mathbf{y}_{3,m_{1}})$. As in \cite{T.Cover},
the decoding error is small if
\begin{equation}\label{equ:rc2}
\alpha tR_{1}\leq \alpha tI(X_{1};Y_{3}|m_{1})+(1-t)R_{0}\leq
\alpha tI(X_{1};Y_{3}|m_{1})+(1-t)I(X_{2};Y_{3}|m_{2}).
\end{equation}
From the cell index $\hat{b}$ and the message index $\hat{d}$
within the cell, the destination can recover the source message.
\end{itemize}
\end{itemize}

Combining \eqref{equ:rc1} and \eqref{equ:rq2}, we have
\begin{equation} \label{eq:pf-R_fb-dmc-bd1}
R<\alpha tI(X_{1};\hat{Y}_{3},Y_{2}|m_{1})+
(1-\alpha)tI(X_{1};Y_{2}|X_{3},m_{3})+(1-t)I(X_{1};Y_{3}|X_{2},m_{2}).
\end{equation}
It follows from \eqref{equ:rc2} and \eqref{equ:rq2} that
\begin{equation} \label{eq:pf-R_fb-dmc-bd2}
R<\alpha tI(X_{1};Y_{3}|m_{1})+(1-t)I(X_{1},X_{2};Y_{3}|m_{2}).
\end{equation}
From \eqref{equ:rhat1} and \eqref{equ:rhat2}, we have the
constraint
\begin{equation} \label{eq:pf-R_fb-dmc-con}
(1-\alpha)tI(X_{3};Y_{2})>\alpha
tI(\hat{Y}_{3};Y_{3}|m_{1})-\alpha tI(\hat{Y}_{3};Y_{2}|m_{1}).
\end{equation}
Thus if~\eqref{eq:pf-R_fb-dmc-bd1},~\eqref{eq:pf-R_fb-dmc-bd2},
and~\eqref{eq:pf-R_fb-dmc-con} are satisfied, there exist a
channel code that makes the decoding error at destination less
than $\epsilon$.
\subsection{Gaussian Channel}
As mentioned in \cite{Kramer.relay}, strong typicality does not
apply to continuous random variables in general, but it does apply
to the Gaussian variables. So the DMC result derived above applies
to the Gaussian $(X_{1},X_{2},X_{3},\hat{Y})$. Since $\hat{Y}_{3}$
is a degraded version of $Y_{3}$, we write $\hat{Y}_{3}=Y_{3}+Z'$
where $Z'$ is Gaussian noise with variance $\sigma_{3}^{2}$
(see~\cite{Gastpar.compress, Madsen.compress} for a similar
analysis).

First, we examine the constraint \eqref{eq:pf-R_fb-dmc-con} under
the Gaussian inputs.
\begin{equation}
I(X_{3};Y_{2}|m_{3})=h(Y_{2})-h(Y_{2}|X_{3})=\frac{1}{2}\log\bigg(\frac{1}{1-\rho_{Y_{2},X_{3}}^{2}}\bigg).
\end{equation}
And
\begin{equation}
\begin{split}
\rho_{Y_{2},X_{3}}^{2}
&=\frac{E^{2}\{(h_{12}X_{1}+h_{23}X_{3}+Z_{2})
X_{3}\}}{Var(Y_{2})Var(X_{3})} \\
&=\frac{\Bigr(h_{12}r_{13}\sqrt{P_{1}^{(3)}}+h_{23}\sqrt{P_{3}^{(3)}}\Bigl)^2}{
h_{12}^{2}P_{1}^{(3)}+h_{23}^{2}P_{3}^{(3)}+\sigma^{2}+
2h_{12}h_{23}r_{13}\sqrt{P_{1}^{(3)}P_{3}^{(3)}}}.
\end{split}
\end{equation}
So
\begin{equation}
I(X_{3};Y_{2}|m_{3})=\frac{1}{2}\log\Bigg(\frac{h_{12}^{2}P_{1}^{(3)}+
h_{23}^{2}P_{3}^{(3)}+\sigma^2+
2h_{12}h_{23}r_{13}\sqrt{P_{1}^{(3)}P_{3}^{(3)}}}{h_{12}^{2}(1-r_{13}^{2})P_{1}^{(3)}+\sigma^2}\Bigg).
\end{equation}

We observe that the correlation coefficient $r_{13}=0$ because
neither the source nor the destination knows the codeword sent by
the other duing the feedback state. Thus, one has
\begin{equation}
I(X_{3};Y_{2}|m_{3})=\frac{1}{2}\log\Bigg(1+
\frac{h_{23}^{2}P_{3}^{(3)}}{h_{12}^{2}P_{1}^{(3)}+\sigma^2}\Bigg)=
C\Bigr(\frac{h_{23}^{2}P_{3}^{(3)}}{h_{12}^{2}P_{1}^{(3)}+\sigma^2}\Bigl).
\end{equation}
Similarly, one has
\begin{equation}
\begin{split}
I(\hat{Y}_{3};Y_{3}|m_{1})-I(\hat{Y}_{3};Y_{2}|m_{1})
&=h(\hat{Y}_{3})
-h(\hat{Y}_{3}|Y_{3})-\frac{1}{2}\log\Bigr(\frac{1}{1-\rho_{\hat{Y}_3Y_2}^{2}}\Bigl) \\
&=h(h_{13}X_{1}+Z_{3}+Z')-h(h_{13}X_{1}+Z_{3}+Z'|h_{13}X_{1}+Z_{3})\\&-
\frac{1}{2}\log\Bigr(\frac{1}{1-\rho_{\hat{Y}_{3}Y_{2}}^{2}}\Bigl) \\
&=\frac{1}{2}\log\Bigr(\frac{h_{13}^{2}P_{1}^{(1)}+\sigma^2+
\sigma_{3}^{2}}{\sigma_{3}^{2}}\Bigl)-\frac{1}{2}\log\Bigr(\frac{1}{1-
\rho_{\hat{Y}_{3}Y_{2}}^{2}}\Bigl)
\end{split}
\end{equation}
where
\begin{equation}
\begin{split}
\rho_{\hat{Y}_{3}Y_{2}}^{2}&=\frac{E^2(\hat{Y}_{3}Y_{2})}{Var(
\hat{Y}_{3})Var(Y_{2})} \\
&=\frac{E^2\{(h_{13}X_{1}+Z_{3}+Z')(h_{12}X_{1}+Z_{2})\}}{Var(\hat{Y}_{3})
Var(Y_{2})} \\
&=\frac{(h_{12}h_{13}P_{1}^{(1)})^2}{(h_{13}^2
P_{1}^{(1)}+\sigma^2+\sigma_{3}^{2})(h_{12}^{2}P_{1}^{(1)}+\sigma^2)}.
\end{split}
\end{equation}
so
\begin{equation}
I(\hat{Y}_{3};Y_{3}|m_{1})-I(\hat{Y}_{3};Y_{2}|m_{1})=
\frac{1}{2}\log\bigg(1+\frac{\sigma^2}{\sigma_{3}^{2}}+
\frac{1}{\sigma_{3}^{2}}\Big(\frac{\sigma^2
h_{13}^{2}P_{1}^{(1)}}{h_{12}^{2}P_{1}^{(1)}+\sigma^2}\Big)\bigg).
\end{equation}
Setting
\begin{equation}
(1-\alpha)tI(X_{3};Y_{2})=\alpha
tI(\hat{Y}_{3};Y_{3}|m_{1})-\alpha tI(\hat{Y}_{3};Y_{2}|m_{1})
\end{equation}
to solve for $\sigma_3^2$
\begin{equation}
\sigma_{3}^{2}=\frac{\sigma^2+\frac{\sigma^2 h_{13}^2
P_{1}^{(1)}}{h_{12}^{2}P_{1}^{(1)}+\sigma^2}}{
(1+\frac{h_{23}^{2}P_{3}^{(3)}}{h_{12}^{2}P_{1}^{(3)}+\sigma^2})^{
\frac{1-\alpha}{\alpha}}-1}.
\end{equation}

Next, we examine the achievable rate expression
\eqref{eq:pf-R_fb-dmc-bd1}.
\begin{equation}
\begin{split}
I(X_{1};\hat{Y}_{3},Y_{2}|m_{1})&=\frac{1}{2}\log\Bigr(1+
\frac{h_{12}^{2}P_{1}^{(1)}}{\sigma^2}+
\frac{h_{13}^{2}P_{1}^{(1)}}{\sigma^2+\sigma_{3}^{2}}\Bigl)
=C\Bigr(\frac{h_{12}^{2}P_{1}^{(1)}}{\sigma^2}+
\frac{h_{13}^{2}P_{1}^{(1)}}{\sigma^2+\sigma_{3}^{2}}\Bigl), \\
I(X_{1};Y_{2}|X_{3},m_{3})&=h(Y_{2}|X_{3},m_{3})-h(Y_{2}|X_{1},X_{3},m_{3}) \\
&=\frac{1}{2}\log\Bigr(1+\frac{h_{12}^{2}P_{1}^{(3)}}{\sigma^2}\Bigl)
=C\Bigr(\frac{h_{12}^{2}P_{1}^{(3)}}{\sigma^2}\Bigl), \\
I(X_{1};Y_{3}|X_{2},m_{2})&=h(Y_{3}|X_{2},m_{2})-
h(Y_{3}|X_{1},X_{2},m_{2}) \\
&=\frac{1}{2}\log\Bigr(1+\frac{(1-r_{12}^{2})h_{13}^{2}P_{1}^{(2)}}{
\sigma^2}\Bigl)=C\Bigr(\frac{(1-r_{12}^{2})h_{13}^{2}P_{1}^{(2)}}{\sigma^2}\Bigl).
\end{split}
\end{equation}
Combining them together, we get
\begin{multline}
\alpha tI(X_{1};\hat{Y}_{3},Y_{2}|m_{1})+
(1-\alpha)tI(X_{1};Y_{2}|X_{3},m_{3})+(1-t)I(X_{1};Y_{3}|X_{2},m_{2}) \\
=\alpha tC\Bigl(\frac{h_{12}^{2}P_{1}^{(1)}}{\sigma^2}+
\frac{h_{13}^{2}P_{1}^{(1)}}{\sigma^2+\sigma_{3}^{2}}\Bigl)+
(1-\alpha)tC\Bigl(\frac{h_{12}^{2}P_{1}^{(3)}}{\sigma^2}\Bigl)+
(1-t)C\Bigl(\frac{(1-r_{12}^{2})h_{13}^{2}P_{1}^{(2)}}{\sigma^2}\Bigl).
\end{multline}
Similarly for \eqref{eq:pf-R_fb-dmc-bd2}, one has
\begin{equation}
\begin{split}
I(X_{1};Y_{3}|m_{1})&=h(h_{13}X_{1}+Z_{3}|m_{1})-h(h_{13}X_{1}+Z_{3}|X_{1},m_{1}) \\
&=\frac{1}{2}\log\Bigl(1+\frac{h_{13}^{2}P_{1}^{(1)}}{\sigma^2}\Bigr)
=C\Bigl(\frac{h_{13}^{2}P_{1}^{(1)}}{\sigma^2}\Bigr), \\
I(X_{1},X_{2};Y_{3}|m_{2})&=h(h_{13}^{2}X_{1}+h_{23}X_{2}+Z_{3}|m_{2})-
h(h_{12}X_{1}+h_{23}X_{2}+Z_{3}|X_{1},X_{2},m_{2}) \\
&=\frac{1}{2}\log\bigg(1+\frac{h_{13}^{2}P_{1}^{(2)}+h_{23}^{2}P_{2}^{2}+
2h_{23}r_{12}\sqrt{P_{1}^{(2)}P_{2}^{(2)}}}{\sigma^2}\bigg) \\
&=C\bigg(\frac{h_{13}^{2}P_{1}^{(2)}+h_{23}^{2}P_{2}^{2}+
2h_{23}r_{12}\sqrt{P_{1}^{(2)}P_{2}^{(2)}}}{\sigma^2}\bigg),
\end{split}
\end{equation}
which gives rise to
\begin{equation}
\begin{split}
 &\alpha tI(X_{1};Y_{3}|m_{1})+(1-t)I(X_{1},X_{2};Y_{3}|m_{2}) \\
&=\frac{\alpha
t}{2}\log\Big(1+\frac{h_{13}^2P_{1}^{(1)}}{\sigma^2}\Big)+
\frac{1-t}{2}\log\Big(1+\frac{h_{13}^{2}P_{1}^{(2)}+h_{23}^{2}P_{2}^{(2)}+
2h_{23}r_{12}\sqrt{P_{1}^{(2)}P_{2}^{(2)}}}{\sigma^2}\Big)\\
&=\frac{\alpha
t}{2}C\Bigg(\frac{h_{13}^2P_{1}^{(1)}}{\sigma^2}\Bigg)+
(1-t)C\Bigg(\frac{h_{13}^{2}P_{1}^{(2)}+h_{23}^{2}P_{2}^{(2)}+
2h_{23}r_{12}\sqrt{P_{1}^{(2)}P_{2}^{(2)}}}{\sigma^2}\Bigg).
\end{split}
\end{equation}
Setting the noise variance $\sigma^2=1$, the proof is complete.

\section{Proof of lemma \ref{lem:up_bds}}\label{appex:up_bds}
We only show the upperbound of $R_{FB}$. The proof for $R_{CF}$ is
similar and thus omitted. Setting shorthand notation
$\Delta=1+\frac{h_{23}^2P_3^{(3)}}{h_{12}^2P_1^{(3)}+1}$, one has
from \eqref{eq:R_fb} that
\begin{equation}
\begin{split}
1+\frac{h_{13}^2P_1^{(1)}}{1+\sigma_3^2}+h_{12}^2P_1^{(1)}
&=\frac{h_{13}^2P_1^{(1)}}{1+\frac{(h_{12}^2+h_{13}^2)P_1^{(1)}+1}{
(h_{12}^2P_1^{(1)}+1)(\Delta^{\frac{1-\alpha}{\alpha}}-1)}}+
(1+h_{12}^2P_1^{(1)}) \\
&=\frac{h_{13}^2P_1^{(1)}(h_{12}^2P_1^{(1)}+1)
(\Delta^{\frac{1-\alpha}{\alpha}}-1)}{\Delta^{\frac{1-\alpha}{\alpha}}
(h_{12}^2P_1^{(1)}+1)+h_{13}^2P_1^{(1)}} + (1+h_{12}^2P_1^{(1)}) \\
&\leq (1+h_{12}^2P_1^{(1)})\Delta^{\frac{1-\alpha}{\alpha}}.
\end{split}
\end{equation}
Hence,
\begin{equation}
\begin{split}
 & \alpha tC\Bigl(\bigl(\frac{h_{13}^2}{1+\sigma_3^2}+h_{12}^2\bigr)
P_1^{(1)}\Bigr) + (1-\alpha)tC\Bigl(h_{12}^2P_1^{(3)}\Bigr) \\
&\leq \alpha tC\Bigl(h_{12}^2P_1^{(1)}\Bigr)+ (1-\alpha)t
C\Bigl(\frac{h_{23}^2P_3^{(3)}}{h_{12}^2P_1^{(3)}+1}\Bigr)+
(1-\alpha)tC\Bigl(h_{12}^2P_1^{(3)}\Bigr) \\
&=\alpha tC\Bigl(h_{12}^2P_1^{(1)}\Bigr)+ (1-\alpha)t
C\Bigl(h_{12}^2P_1^{(3)}+h_{23}^2P_3^{(3)}\Bigr)
\end{split}
\end{equation}
which proves \eqref{eq:R_fb_ub}.

\section{Proof of theorem \ref{thm:relaycomph}}\label{appex:relaycomph}
In view of $R_{DF}$ in \eqref{eq:R_df}, $h_{12}^2\leq h_{13}^2$
implies that
\begin{equation}
R_{DF}\leq tC\Bigl(h_{12}^2P_1^{(1)}\Bigr)+(1-t)C\Bigl(
(1-r_{12}^2)h_{13}^2P_1^{(2)}\Bigr)\leq
C\Bigl(h_{13}^2P\Bigr)=R_{ro}
\end{equation}
where we have used the total power constraint
\eqref{equ:relaypow}. To prove 2), consider the upperbound for
$R_{CF}$ in \eqref{eq:R_cf_ub}. Given the total power constraint
$P_1^{(2)}+P_2^{(2)}\leq P$, it is easy to verify that
$h_{13}^2P_1^{(2)}+h_{23}^2P_2^{(2)}\leq\max\{h_{13}^2,h_{23}^2\}P$.
Therefore, the condition $h_{23}^2\leq h_{13}^2$ implies that
\begin{equation}
R_{CF}\leq tC\Bigl(h_{13}^2P_1^{(1)}\Bigr)+(1-t)C\Bigl(h_{13}^2
P_1^{(2)}+h_{23}^2P_2^{(2)}\Bigr)\leq C\Bigl(h_{13}^2P\Bigr)
=R_{ro}.
\end{equation}
The last statement of the theorem can be shown in a similar
fashion using the $R_{FB}$ upperbound in \eqref{eq:R_fb_ub}.

\section{Proof of theorem \ref{thm:relaylowP}}\label{appex:relaylowP}

Since $C(h_{12}^2P)\geq C(h_{13}^2P)$ by $h_{12}^2\geq h_{13}^2$,
the two line segments in the $R_{DF}$ expression intersect at some
optimal $t^* \in(0,1)$ (see Fig.~\ref{fig:linebd}). The
corresponding rate is given by
\begin{equation}
R_{DF}=\frac{C\Bigl(f_1(\theta,r_{12},h_{13},h_{23})P\Bigr)
C\Bigl(h_{12}^2P\Bigr)-C\Bigl(f_2(\theta,r_{12},h_{13})P\Bigr)
C\Bigl(h_{13}^2P\Bigr)}{C\Bigl(f_1(\theta,r_{12},h_{13},h_{23})P\Bigr)
+ C\Bigl(h_{12}^2P\Bigr)-C\Bigl(f_2(\theta,r_{12},h_{13})P\Bigr) -
C\Bigl(h_{13}^2P\Bigr) }
\end{equation}
where we have set $P_1^{(2)}=P\cos^2\theta$ and
$P_2^{(2)}=P\sin^2\theta$ according to the total power constraint.
Taking $P\rightarrow 0$, the Taylor expansion is sufficient to
establish \eqref{eq:S_df}. To prove the lowerbound in
\eqref{eq:S_df}, note that $f_1(\theta,r_{12},h_{13},h_{23})\leq
(h_{13}^2+h_{23}^2)$ with equality when $r_{12}=1$ and
$\tan(\theta)=\frac{h_{23}}{h_{13}}$, which, together with
$f_2(\theta,r_{12},h_{13})\leq h_{13}^2$, also proves the
upperbound of $S_{DF}$ in \eqref{eq:S_df}.

On the other hand, as $P\rightarrow 0$, it is seen from
\eqref{eq:R_cf} that $\sigma_2^2\rightarrow \infty$, thus showing
$S_{CF}\leq h_{13}^2$. The similar behavior holds for the feedback
scheme, that is, $\sigma_3^2\rightarrow\infty$ as $P\rightarrow
0$, in which case $S_{FB}\leq S_{DF}$ with the optimal $\alpha$
approaches $1$.

\section{Proof of theorem \ref{thm:relayhighP}}\label{appex:relayhighP}
The results for $G_{DF}$ and $G_{CF}$ follows from direct
computation of large $P$ limit. We only show the last statement
concerning the feedback scheme. As in the case of decode-forward,
the line-crossing point gives the optimal $t$ and the associated
rate $R_{FB}$ is given by
\begin{equation}
\frac{C\Bigl(f_1(\theta,r_{12},h_{13},h_{23})P\Bigr) A-
C\Bigl(f_2(\theta,r_{12},h_{13})P\Bigr)\alpha
C\Big(h_{13}^2P\Bigr)}{
C\Bigl(f_1(\theta,r_{12},h_{13},h_{23})P\Bigr)+ A -
C\Bigl(f_2(\theta,r_{12},h_{13})P\Bigr) - \alpha
C\Big(h_{13}^2P\Bigr) },
\end{equation}
in which
\begin{equation}
A=\alpha C\Bigl(\bigl(\frac{h_{13}^2}{1+\sigma_3^2}+h_{12}^2\bigr)
P\Bigr)+(1-\alpha)C\Bigl(h_{12}^2P\cos^2\psi\Bigr),
\end{equation}
where we set $P_1^{(3)}=P\cos^2\psi$ and $P_3^{(3)}=P\sin^2\psi$.
Taking $P\rightarrow\infty$,
\begin{equation}
\sigma_3^2\rightarrow \frac{h_{12}^2+h_{13}^2}{h_{12}^2\Bigl[
(1+\frac{h_{23}^2}{h_{12}^2}\tan^2\psi)^{\frac{1-\alpha}{\alpha}}
-1\Bigr]} \quad (=\sigma_3^2(\infty)).
\end{equation}
Denote $f_3(\psi,\alpha,h_{13},h_{12},h_{23})=\alpha\log\Bigl(
\frac{h_{13}^2}{1+\sigma_3^2(\infty)}+h_{12}^2\Bigr) + (1-\alpha)
\log h_{12}^2\cos^2\psi $, one has
\begin{equation}
R_{FB}\sim \frac{1}{2}\log P + \frac{\frac{1}{4}(1-\alpha)\log
f_2\cdot \log P + \frac{1}{4}\Bigl[\log f_1\cdot\log f_3 -
\alpha\log f_2\cdot \log h_{13}^2\Bigr]}{\frac{1-\alpha}{2}\log P
+ \frac{1}{2}\Bigl[ \log f_1 + \log f_3 - \log f_2 -\alpha\log
h_{13}^2\Bigr]}.
\end{equation}
It follows that if $\alpha<1$
\begin{equation}
G_{FB}\leq \log f_2 \leq \log h_{13}^2 \quad (\text{relay-off}),
\end{equation}
which forces $\alpha=1$, that is, $G_{FB}=G_{DF}$.

\section{Proof of theorem \ref{Them:multicastlim}}\label{appex:multicast}
Here we only prove the part 1) of this theorem. Parts 2) - 5)
follow the same lines as the corresponding results in the relay
case.

To prove part 1), it suffices to show the statement for
$\alpha=1$. The capacity of the multicast channel without
cooperation is given by $R_{non-coop}=C(\min
\{h_{13}^{2},h_{12}^2\}P)$. With the assumption that
$h_{12}^2>h_{13}^2$, we have $R_{non-coop}=C(h_{13}^2 P)$.

Note that the rate expression of \eqref{equ:multicastgau} admits
the same line-crossing interpretation as in the relay case. Thus,
the intersection determines the optimal rate point. Equate the two
terms
\begin{equation}
tC\bigl(h_{12}^2P_1^{(1)}\bigr)=tC\bigl(h_{13}^2P_1^{(1)}\bigr)+
(1-t)C((h_{13}^{2}+h_{23}^{2})P)
\end{equation}
to solve
\begin{equation}
t^*=\frac{C((h_{13}^{2}+h_{23}^{2})P)}{C((h_{13}^{2}+h_{23}^{2})P)+
C(h_{12}^{2}P)-C(h_{13}^{2}P)},
\end{equation}
which gives the corresponding rate
\begin{equation}
R_{DF}=\frac{C\bigl((h_{13}^2+h_{23}^2)P\bigr)C\bigl(h_{12}^2P\bigr)}{
C\bigl((h_{13}^2+h_{23}^2)P\bigr)+C\bigl(h_{12}^2P\bigr)-
C\bigl(h_{13}^2P\bigr)}.
\end{equation}
Therefore, using $h_{12}^2>h_{13}^2$, one has
\begin{equation}
R_{DF}-R_{non-coop}=(1-t^*)\bigl(C((h_{13}^2+h_{23}^2)P)-
C(h_{13}^2P)\bigr)>0
\end{equation}
which proves the theorem.

\section{Proof of lemma \ref{lemma:multicastsideinfor}}\label{appex:multiside}
As Appendix~\ref{appex:fb}, we first prove the result for DMC
case, then apply the result to the Gaussian channel.
\subsection{Source coding}
Randomly bin all the sequence $S_{1}^K$ into
$2^{K(H(S_{1}|S_{2})+\epsilon)}$ bins by independently generating
an index $w$ uniformly distributed on
$\{1,2,...,2^{K(H(S_{1}|S_{2})+\epsilon)}\}$. Let $f_{s1}$ be the
mapping function, such that $w=f_{s1}(s_{1}^K)$. Independently
generate another bin index $b$ for every sequence $S_{1}^K$ by
picking $b$ uniformly from $\{1,2,...,2^{KR}\}$. Let $B(b)$ be the
set of all sequences $S_{1}^K$ allocated to bin $b$. Thus, every
source sequence is associated with two bin indexes $\{w,b\}$.

\subsection{Channel coding}
\subsubsection{Random code generation}

\begin{itemize}
\item At state $m_{1}$, generate $2^{K(H(S_{1}|S_{2})+\epsilon)}$
i.i.d.\ length-$N_{1}$ sequence $\mathbf{x}_{1,m_{1}}$, each with
probability
$p(\mathbf{x}_{1,m_{1}})=\prod\limits_{j=1}^{N_{1}}p(x_{1}|m_{1})$,
in which $p(x_{1}|m_{1})$ is the input distribution that maximizes
$I(X_{1};Y_{2})$. Assign every bin index $w$ to one sequence
$\mathbf{x}_{1,m_{1}}(w),w\in [1,2^{K(H(S_{1}|S_{2})+\epsilon)}]$.

\item At state $m_{2}$, randomly generate $2^{KR}$ i.i.d.\
length-$N_{0}$ $\mathbf{x}_{2,m_{2}}$ at node-$2$, each with
probability
$p(\mathbf{x}_{2,m_{2}})=\prod\limits_{j=1}^{N_{0}}p(x_{2j}|m_{2})$.
Generate $2^{KR}$ i.i.d.\ length-$N_{0}$ $\mathbf{x}_{1,m_{2}}$ at
node-$1$, each with probability
$p(\mathbf{x}_{1,m_{2}})=\prod\limits_{j=1}^{N_{0}}p(x_{1j}|m_{2})$
, in which $p(x_{1}|m_{2})=\sum \limits_{x_{2}}
p(x_{1},x_{2}|m_{2})$. And $p(x_{1},x_{2}|m_{2})$is the input
distribution that maximizes $I(X_{1},X_{2};Y_{3})$. Associate
every bin index $b$ to one sequence pair
$\{\mathbf{x}_{1,m_{2}}(b),\mathbf{x}_{2,m_{2}}(b)\}$.

\end{itemize}

\subsubsection{Coding}
Suppose we want to send source sequence $s_{1}^K(i)$ at block $i$,
and $w(i)=f_{s1}(s_{1}^K(i))$, $s_{1}^K(i)\in B(b(i))$. For
brevity of notation, we drop block index $i$ in the following.
\begin{itemize}

\item State $m_{1}$:\newline Node-$1$ sends
$\mathbf{x}_{1,m_{1}}(w)$.

\item State $m_{2}$:
\begin{itemize}

\item Node-$1$ knows $s_{1}^K$ is in $b$, so it sends
$\mathbf{x}_{1,m_{2}}(b)$.

\item At the end of state $m_{1}$, node-$2$ gets an estimation of
$\hat{s}_{21}^K$ (details will be given in the following), and
suppose $\hat{s}_{21}^K$ is in bin $\hat{b}$. Then in state
$m_{2}$ nodes-$2$ sends the corresponding
$\mathbf{x}_{2,m_{2}}(\hat{b})$.

\end{itemize}
\end{itemize}

\subsection{Decoding}
At the end of state $m_{1}$:
\begin{itemize}

\item At node-$2$:\newline At first, node-$2$ looks for the one
and only one $\hat{w}$ such that
$\{\mathbf{x}_{1,m_{1}}(\hat{w}),\mathbf{y}_{2,m_{1}}\}$ are
jointly typical. Then node-$2$ searches in the bin indexed by
$\hat{w}$ for source sequence $\hat{s}_{21}^K$ such that
$\{\hat{s}_{21}^K,s_{2}^K\}$ are jointly typical. If it finds only
one such sequence, it declares it has received $\hat{s}_{21}^K$,
otherwise declares an error.

\item At node-$3$:\newline Node-$3$ calculates a list
$\ell(\mathbf{y}_{3,m_{1}})$, such that $w'\in
\ell(\mathbf{y}_{3,m_{1}})$ if
$\{\mathbf{x}_{1,m_{1}}(w'),\mathbf{y}_{3,m_{1}}\}$ are jointly
typical.
\end{itemize}

At the end of state $m_{2}$, only node-$3$ needs to decode:
\begin{itemize}
\item Step 1:
\newline node-$3$ declares it receives $\hat{\hat{b}}$, if $\hat{\hat{b}}$
is the one and only one index such that
$\{\mathbf{x}_{1,m_{2}}(\hat{\hat{b}}),\mathbf{x}_{2,m_{2}}(\hat{\hat{b}}),\mathbf{y}_{3,m_{2}}\}$
are jointly typical.

\item Step 2:
\newline node-$3$ searches in the bin $B(\hat{\hat{b}})$ for the one and
only one source sequence $\hat{s}_{31}^K$, such that
$\{\hat{s}_{31}^K,s_{3}^K\}$ are jointly typical and
$f_{s1}(\hat{s}_{31}^K)\in\ell(\mathbf{y}_{3,m_{1}})$. If it finds
such a unique one, it declares that $\hat{s}_{31}^K$ is the source
sequence. Otherwise it declares an error.

\end{itemize}

\subsection{Calculation of Probability of Error}
\subsubsection{Node-$2$}
For node-$2$ there are following error events:
\begin{eqnarray}
E_{0}&=&\{(s_{1}^K,s_{2}^K)\notin A_{\epsilon}^K\},\\
E_{1}&=&\{\hat{w}\neq w\},\\
E_{2}&=&\{\exists s_{1}^{'K}:s_{1}^{'K}\neq s_{1}^K,
f_{s1}(s_{1}^{'K})=w\;\text{and}\;(s_{1}^{'K},s_{2}^K)\in
A_{\epsilon}^K\}.
\end{eqnarray} And
\begin{equation}
P_{e}^{N_{1},K}=P(E_{0}\cup E_{1}\cup E_{2}) \leq
P(E_{0})+P(E_{1}|E_{0}^c)+P(E_{2}|E_{0}^c,E_{1}^c).
\end{equation}
When $K$ is sufficiently large, using the AEP,
$P(E_{0})\rightarrow 0$. Now consider $P(E_{1}|E_{0}^c)$, if
channel code rate is less than the capacity, receiver will decode
channel code with error probability less than $\epsilon$. Here,
there are $2^{K(H(S_{1}|S_{2})+\epsilon)}$ code words, and channel
code length is $N_{1}$, then the rate of channel code is
$\frac{K(H(S_{1}|S_{2})+\epsilon)}{N_{1}}$. Thus for sufficently
large $N_{1}$ and $K$, $P(E_{1}|E_{0}^c)\leq \epsilon)$ if
\begin{equation}
\frac{K(H(S_{1}|S_{2})+\epsilon)}{N_{1}}<\max\limits_{p(x_{1})}
I(X_{1};Y_{2}|m_{1})=C_{2},
\end{equation}
which is the same as:
\begin{equation}\label{equ:N1condi}
H(S_{1}|S_{2})+\epsilon<\tau_{1}C_{2}.
\end{equation}
Because source code rate is $H(S_{1}|S_{2})+\epsilon$, using the
same argument as \cite{T.Cover1}, one can get
$P(E_{2}|E_{0}^c,E_{1}^c)<\epsilon$, if $K$ is sufficiently large.
So if (\ref{equ:N1condi}) is satisfied, and $N_{1},K$ are
sufficiently large, there exists a source-channel code that make
the error probability at node-$2$
\begin{equation}
P_{e}^{N_{1},K}=P(\hat{s}_{21}^K\neq s_{1}^K)\leq 3\epsilon.
\end{equation}

\subsubsection{Node-$3$} For node-$3$, there are following error
events:
\begin{eqnarray}E_{0}&=&\{(s_{1}^K,s_{3}^K)\notin A_{\epsilon}^K\},\\
E_{1}&=&\{\text{node 2 can not decode successfully}\},\\
E_{2}&=&\{\hat{\hat{b}}\neq b\},\\
E_{3}&=&\{\exists s_{1}^{'K}: s_{1}^{'K}\neq s_{1}^K,
f_{s1}(s_{1}^{'K})\in \ell(\mathbf{y}_{3}|m_{1}),s_{1}^{'K}\in
B(\hat{\hat{b}}),(s_{1}^{'K},s_{3}^K)\in A_{\epsilon}^K\}.
\end{eqnarray}
\begin{equation}
P_{e}^{N,K}=P(E_{0}\cup E_{1}\cup E_{2}\cup E_{3})\leq
P(E_{0})+P(E_{1})+P(E_{2}|E_{0}^c,E_{1}^c)+P(E_{3}|E_{0}^c,E_{1}^c,E_{2}^c).
\end{equation}
When $K$ is sufficiently large, $P(E_{0})\rightarrow 0$. And if
\eqref{equ:N1condi} is satisfied, $P(E_{1})\leq 3\epsilon$. Now
consider $P(E_{2}|E_{0}^c,E_{1}^c)$, the channel code rate is
$\frac{KR}{N_{0}}$. So, $P(E_{2}|E_{0}^c,E_{1}^c)\leq \epsilon$
for sufficiently large $N_{0}$, if
\begin{equation}
\frac{KR}{N_{0}}\leq \max \limits_{p(x_{1},x_{2})}
I(X_{1},X_{2};Y_{3})=C_{(1,2)-3},
\end{equation}
that is,
\begin{equation}\label{eq:m_s_r}
R\leq \tau_{0} C_{(1,2)-3}.
\end{equation}

Now consider $P(E_{3}|E_{0}^c,E_{1}^c,E_{2}^c)$:
\begin{equation}
\begin{split}
P(E_{3}|E_{0}^c,E_{1}^c,E_{2}^c) &=P(\exists
s_{1}^{'K}:s_{1}^{'K}\neq S_{1}^K, f_{s1}(s_{1}^{'K})\in
\ell(\mathbf{Y}_{3}|m_{1}),s_{1}^{'K}\in B(b),
(s_{1}^{'K},S_{3}^K)\in A_{\epsilon}^K)\\
&=\sum \limits_{(s_{1}^K,s_{3}^K)}p(s_{1}^K,s_{3}^K)P(\exists
s_{1}^{'K}\neq s_{1}^K, f_{s1}(s_{1}^{'K})\in
\ell(\mathbf{y}_{3}|m_{1}), s_{1}^{'K}\in B(b),
(s_{1}^{'K},s_{3}^K)\in A_{\epsilon}^K)\\
&\leq \sum \limits_{(s_{1}^K,s_{3}^K)}p(s_{1}^K,s_{3}^K)\sum
\limits_{s_{1}^{'K}\neq s_{1}^K
\;\text{and}\;(s_{1}^{'K},s_{3}^K)\in
A_{\epsilon}^K}P(f_{s1}(s_{1}^{'K})\in
\ell(\mathbf{y}_{3}|m_{1}), s_{1}'^K\in B(b))\\
&=\sum \limits_{(s_{1}^K,s_{3}^K)}p(s_{1}^K,s_{3}^K)\sum
\limits_{s_{1}^{'K}\neq s_{1}^K
\;\text{and}\;(s_{1}^{'K},s_{3}^K)\in
A_{\epsilon}^K}P(f_{s1}(s_{1}^{'K})\in
\ell(\mathbf{y}_{3}|m_{1}))P(s_{1}'^K\in B(b))\\
&\leq \sum
\limits_{(s_{1}^K,s_{3}^K)}p(s_{1}^K,s_{3}^K)2^{-K(H(S_{1}|S_{2})+\epsilon)}
\|\ell(\mathbf{y}_{3}|m_{1})\|2^{-KR}\|A_{\epsilon}(S_{1}^K|s_{3}^K)\|\\
&\leq
2^{-K(H(S_{1}|S_{2})+\epsilon)}E\{\|\ell(\mathbf{y}_{3}|m_{1})
\|\}2^{-KR}2^{K(H(S_{1}|S_{3})+\epsilon)}.
\end{split}
\end{equation}
Follow the same steps in the~\cite{T.Cover}, one has
$E\{\|\ell(\mathbf{y}_{3}|m_{1})\|\}\leq
1+2^{K(H(S_{1}|S_{2})+\epsilon)}2^{-N_{1}(I(X_{1};Y_{3}|m_{1})-7\epsilon)}$.
So
\begin{equation}
\begin{split}
P(E_{3}|E_{0}^c,E_{1}^c,E_{2}^c) &\leq
2^{-K(H(S_{1}|S_{2})+\epsilon)}(1+2^{K(H(S_{1}|S_{2})+\epsilon)}2^{-N_{1}(I(X_{1};Y_{3}|m_{1})-7\epsilon)})
2^{-KR}2^{K(H(S_{1}|S_{3})+\epsilon)}
\\
&=2^{-K\{R-(H(S_{1}|S_{3})+\epsilon)+(H(S_{1}|S_{2})+\epsilon)\}}+
2^{-K\{R+\frac{N_{1}}{K}(I(X_{1};Y_{3}|m_{1})-7\epsilon)-(H(S_{1}|S_{3})+\epsilon)\}}
\end{split}
\end{equation}

So if \begin{equation}
R>H(S_{1}|S_{3})+\epsilon-(H(S_{1}|S_{2})+\epsilon),
\end{equation}
and
\begin{equation}
R>H(S_{1}|S_{3})+\epsilon-\frac{N_{1}}{K}(I(X_{1};Y_{3}|m_{1})-7\epsilon)>H(S_{1}|S_{3})+
\epsilon-\tau_{1}I(X_{1};Y_{3}|m_{1}),
\end{equation}
and $K$ is sufficiently large,
$P(E_{3}|E_{0}^c,E_{1}^c,E_{2}^c)\leq \epsilon$. Together with
\eqref{equ:N1condi} and \eqref{eq:m_s_r}, one can get
\begin{equation}\label{eq:ap_tau0}
H(S_{1}|S_{3})+\epsilon-\frac{\min
\{I(X_{1};Y_{3}|m_{1})-7\epsilon,C_{2}\}(H(S_{1}|S_{2})+\epsilon)}{C_{2}}\leq
\tau_{0}C_{(1,2)-3}.
\end{equation}
Thus, if both~\eqref{eq:ap_tau0} and~\eqref{equ:N1condi} are
satisfied, there exists a source-channel code that makes the error
probability at node-$3$ $P_{e}^{N,K}<6\epsilon$.

Next step is to apply the result to the Gaussian channel. In this
case, we have
\begin{equation}\label{equ:gau}
\begin{split}
C_{2}&=C(h_{12}^2P),\\
I(X_{1};Y_{3}|m_{1})&=C(h_{13}^2P),\\
C_{(1,2)-3}&=C((h_{13}^2+h_{23}^2)P).
\end{split}
\end{equation}
Inserting~\eqref{equ:gau} to~\eqref{eq:ap_tau0}
and~\eqref{equ:N1condi} completes the proof.

\section{Proof of Theorem~\ref{them:multisideener}}
\label{appex:multisideener} Part 1) and 2) of this theorem follow
straightforward limit calculation, we only prove part 3).

The assumption $\tau_{ex,2}<\tau_{ex,3}$ becomes
$\frac{H(S_{1}|S_{2})}{h_{12}^2}<\frac{H(S_{1}|S_{3})}{h_{13}^2}$
when $P\rightarrow 0$. Under this assumption, there are two
different cases corresponding to different cost function for the
benchmark scheme: $H(S_{1}|S_{2})>H(S_{1}|S_{3})$ and
$H(S_{1}|S_{2})<H(S_{1}|S_{2})$.
\newline When
$H(S_{1}|S_{2})>H(S_{1}|S_{3})$, in which case $h_{12}^2>h_{13}^2$
and
\begin{equation}
E_{1,m}=\frac{2}{\log
e}\bigg(\frac{H(S_{1}|S_{2})}{h_{12}^2}+\big(\frac{1}{h_{13}^2}
-\frac{1}{h_{12}^2}\big)^{+}H(S_{1}|S_{3})\bigg).
\end{equation}
\begin{equation}
\begin{split}
E_{2,m}&=\frac{2}{\log
e}\bigg(\frac{H(S_{1}|S_{2})}{h_{12}^2}+\frac{h_{12}^2H(S_{1}|S_{3})
-h_{13}^2H(S_{1}|S_{2})}{(h_{13}^2+h_{23}^2)h_{12}^2}\bigg)\\
&< \frac{2}{\log
e}\bigg(\frac{H(S_{1}|S_{2})}{h_{12}^2}+\frac{h_{12}^2H(S_{1}|S_{3})-
h_{13}^2H(S_{1}|S_{3})}{(h_{13}^2+h_{23}^2)h_{12}^2}\bigg)\\
&< \frac{2}{\log
e}\bigg(\frac{H(S_{1}|S_{2})}{h_{12}^2}+\frac{h_{12}^2H(S_{1}|S_{3})-
h_{13}^2H(S_{1}|S_{3})}{h_{13}^2h_{12}^2}\bigg)\\
&\leq \frac{2}{\log
e}\bigg(\frac{H(S_{1}|S_{2})}{h_{12}^2}+\big(\frac{1}{h_{13}^2}-
\frac{1}{h_{12}^2}\big)^{+}H(S_{1}|S_{3})\bigg)\\
&=E_{1,m}.
\end{split}
\end{equation}
When $H(S_{1}|S_{2})>H(S_{1}|S_{3})$,
\begin{equation}
E_{1,m}=\frac{2}{\log
e}\bigg(\frac{H(S_{1}|S_{3})}{h_{13}^2}+\big(\frac{1}{h_{12}^2}
-\frac{1}{h_{13}^2}\big)^{+}H(S_{1}|S_{2})\bigg),
\end{equation}
so
\begin{equation}
\begin{split}
E_{2,m}&=\frac{2}{\log
e}\bigg(\frac{H(S_{1}|S_{2})}{h_{12}^2}+\frac{h_{12}^2H(S_{1}|S_{3})
-\min\{h_{13}^2,h_{12}^2\}H(S_{1}|S_{2})}{(h_{13}^2+h_{23}^2)h_{12}^2}\bigg)\\
&<\frac{2}{\log
e}\bigg(\frac{H(S_{1}|S_{2})}{h_{12}^{2}}+\frac{h_{12}^2H(S_{1}|S_{3})
-\min\{h_{13}^2,h_{12}^2\}H(S_{1}|S_{2})}{h_{13}^2h_{12}^2}\bigg)\\
&=\frac{2}{\log
e}\bigg(\frac{H(S_{1}|S_{2})}{h_{12}^{2}}+\frac{H(S_{1}|S_{3})}{h_{13}^2}-
\min\{\frac{1}{h_{13}^2},\frac{1}{h_{12}^2}\}H(S_{1}|S_{2})\bigg)\\
&=\frac{2}{\log
e}\bigg(\frac{H(S_{1}|S_{3})}{h_{13}^{2}}+(\frac{1}{h_{12}^2}
-\frac{1}{h_{13}^2})^{+}H(S_{1}|S_{2})\bigg)\\
&=E_{1,m}.
\end{split}
\end{equation}

\section{Proof of Theorem~\ref{thm:conference}}\label{appex:conference}

For part 1) of this theorem, without loss of generality, we only
prove the case when $h_{23}^2\rightarrow \infty$. In this case,
$\lim \limits_{h_{23}^2\rightarrow \infty}\tau_{2,gen}= \lim
\limits_{h_{23}^2\rightarrow \infty}\tau_{3,gen}=0$,
$\tau_{gen}=\tau_{1,gen}$. In the following, we will show that the
genie-aided bound could be achieved using the following multicast
order $2\rightarrow3\rightarrow1$.

When node-$2$ multicasts $S_{2}^K$ to both node-$3$ and node-$1$
using the proposed cooperative multicast with side-information
scheme, from Lemma~\ref{lem:multicastsidefb} we know it requires
\begin{equation}
\tau_{2-(3,1)}=\frac{H(S_{2}|S_{3})}{R_{CFr1d3}(\alpha)}
+\frac{H(S_{2}|S_{1})-\alpha
H(S_{2}|S_{3})\frac{\min\{I(X_{2};Y_{1}),I(X_{2};\hat{Y}_{1},Y_{3})\}}
{R_{CFr1d3}(\alpha)}}{C((h_{12}^2+h_{13}^2)P)}.\nonumber
\end{equation}
$R_{CFr1d3}$ means the achievable rate of the following relay
channel using Compress-Forward scheme: node-$2$ is the source,
node-$1$ acts as relay that spends $1-\alpha$ part of the time in
helping destination using CF scheme, and node-$3$ acts as the
destination.

Next consider node-$3$ multicasts $S_{3}^K$ to both node-$1$,
node-$2$. At this time, node-$1$ already has $S_{1},\;S_{2}$, thus
this step requires
\begin{equation}
\tau_{3-(2,1)}=\frac{H(S_{3}|S_{2})}{R_{CFr1d2}(\alpha)}
+\frac{H(S_{3}|S_{1},S_{2})-\alpha
H(S_{2}|S_{3})\frac{\min\{I(X_{3};Y_{1}),I(X_{3};\hat{Y}_{1},Y_{2})\}}
{R_{CFr1d2}(\alpha)}}{C((h_{12}^2+h_{13}^2)P)}.\nonumber
\end{equation}

Final step, node-$1$ multicasts $H(S_{1}|S_{2},S_{3})$ to both
node-$2$, node-$3$ using the greedy multicast scheme developed in
the multicast section, this step requires $\tau_{1-(
2,3)}=\frac{H(S_{1}|S_{2},S_{3})}{R_{g}}$.

Thus, the total bandwidth expansion factor of this scheme is
\begin{equation}
\tau=\tau_{2-(3,1)}+\tau_{3-(1,2)}+\tau_{1-(2,3)}.
\end{equation}
Based on the results on the relay channel and multicast channel
$h_{23}^2\rightarrow \infty$, $R_{CFr1d3}\rightarrow\infty$,
$R_{CFr1d2}\rightarrow\infty$, $\lim \limits_{h_{23}^2\rightarrow
\infty} R_{g} =C((h_{12}^2+h_{13}^2)P)$. Then
\begin{equation}
\begin{split}
\lim \limits_{h_{23}\rightarrow \infty}\tau &=
\frac{H(S_{2}|S_{1})+H(S_{3}|S_{1},S_{2})
+H(S_{1}|S_{2},S_{3})}{C((h_{12}^2+h_{13}^2)P)}\\
&=\frac{H(S_{2},S_{3}|S_{1})+H(S_{1}|S_{2},S_{3})}{C((h_{12}^2+h_{13}^2)P)}
=\tau_{1,gen}=\tau_{gen}.
\end{split}
\end{equation}

To prove the second part of this theorem, without loss of
generality, suppose $1\rightarrow2\rightarrow3$ is the optimal
multicast order for the scheme that uses broadcast with degraded
information set. Then, just use the same order for the cooperative
source-channel coding scheme based multicast with
side-information. Theorem \ref{them:multisideener}
 shows that at every multicast step, the cooperative source-channel coding scheme
 outperforms the broadcast with degraded information set. Thus even with this
 not necessarily optimal order, the cooperative source-channel coding scheme outperforms
 the scheme that uses broadcast with degraded information set with optimal order.
\newpage

\begin{figure}[htb]
\begin{center}
\includegraphics[width=0.4 \textwidth]{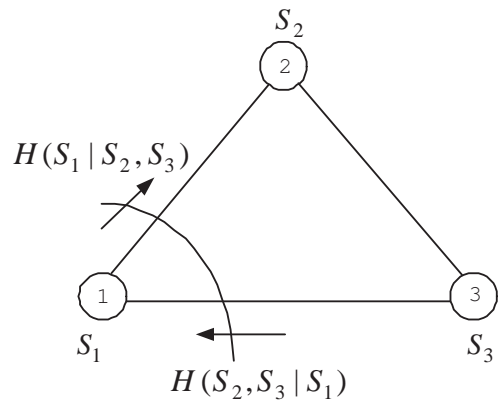}
\caption{The genie-aided bound in node-$1$, in which node-$2$ and
node-$3$ can fully cooperate with each other.}
\label{fig:conferencechannelupper}
\end{center}
\end{figure}


\end{document}